\RequirePackage{fix-cm}
\documentclass[twocolumn,smallcondensed,final]{svjour3} 
\smartqed  
\journalname{Cluster Computing}
\usepackage[export]{adjustbox}

\usepackage{multirow} 
\usepackage{algpseudocode}
\usepackage{algorithm}
\usepackage{rotating}

\usepackage{kantlipsum} 
\usepackage{commath}
\usepackage{mathtools}  
\usepackage{xr-hyper} 
\usepackage{enumitem}
\usepackage{hhline}
\usepackage[table,xcdraw]{xcolor}
 \usepackage{graphicx,booktabs}
 \usepackage{longtable}
\usepackage[pdfusetitle, pdfauthor={Michael Shell, My institution}]{hyperref}

\usepackage{epstopdf}
\usepackage{wrapfig}
\usepackage{latexsym}
\usepackage{amssymb}
\usepackage{amsfonts}
\usepackage{graphicx}
\usepackage{latexsym}
\usepackage{booktabs}
\usepackage[style=base]{caption}
\usepackage{subcaption} 
\usepackage{breqn}

\algnewcommand\algorithmicinput{\textbf{INPUT:}}
\algnewcommand\INPUT{\item[\algorithmicinput]}
\algnewcommand\algorithmicoutput{\textbf{OUTPUT:}}
\algnewcommand\OUTPUT{\item[\algorithmicoutput]}
\usepackage[table]{xcolor}
\usepackage{tikz}
\usetikzlibrary{fit} 

\usepackage{color}
\colorlet{BLUE}{blue}
\usepackage{colortbl}
\definecolor{LightCyan}{RGB}{155, 227, 247}

\newif\ifcommentson

\newif\ifextended
\newif\ifshortver

\newcommand{\optional}[1]{\ignorespaces}

\newif\ifrevisionactive
\newif\ifshowdeleted
\revisionactivetrue

\begin{document}



\title{Can Machine Learning Model with Static Features be Fooled: an Adversarial Machine Learning Approach}

\titlerunning{Can Machine Learning Model with Static Features be Fooled: an Adversarial Machine Learning Approach}
	\author{Rahim~Taheri \and Reza~Javidan \and Mohammad~Shojafar\text{*} \and Vinod~P \and Mauro Conti}
\institute
    {
    R. Taheri and R. Javidan \at Computer Engineering and IT Department, \\Shiraz University of Technology, Shiraz, Iran\\\email{\{r.taheri, javidan\}sutech.ac.ir}
	\and 
	M. Shojafar, Vinod P., and M. Conti \at Department of Mathematics,\\ University of Padua, Padua, Italy\\
	\email{pvinod21@gmail.com, mohammad.shojafar@unipd.it; m.shojafar@ieee.org, conti@math.unipd.it}
	}
\date{Received: 08 October 2019 / Revised: 22 February 2020/ Accepted: 29 February 2020}
\maketitle



\begin{abstract}
The widespread adoption of smartphones dramatically increases the risk of attacks and the spread of mobile malware, especially on the Android platform. Machine learning-based solutions have been already used as a tool to supersede signature-based anti-malware systems. However, malware authors leverage features from malicious and legitimate samples to estimate statistical difference in-order to create adversarial examples. Hence, to evaluate the vulnerability of machine learning algorithms in malware detection, we propose five different attack scenarios to perturb malicious applications (apps). By doing this, the classification algorithm inappropriately fits the discriminant function on the set of data points, eventually yielding a higher misclassification rate. Further, to distinguish the adversarial examples from benign samples, we propose two defense mechanisms to counter attacks. To validate our attacks and solutions, we test our model on three different benchmark datasets. We also test our methods using various classifier algorithms and compare them with the state-of-the-art data poisoning method using the Jacobian matrix. Promising results show that generated adversarial samples can evade detection with a very high probability. Additionally, evasive variants generated by our attack models when used to harden the developed anti-malware system improves the detection rate up to 50\% when using the Generative Adversarial Network (GAN) method.
\end{abstract}
\keywords{
Adversarial machine learning \and Android malware detection \and poison attacks \and generative adversarial network \and Jacobian algorithm.} 

\section{Introduction}\label{sez:1}


Nowadays using the Android application is very popular on mobile platforms. Every Android application has a Jar-like APK format and is an archive file which contains Android manifest and \textit{Classes.dex} files. Information about the structure of the Apps holds in the manifest file and each part is responsible for certain actions. For instance, the requested permissions must be accepted by the users for successful installation of applications. The manifest file contains a list of hardware components and permissions required by each application. 
Furthermore, there are environment settings in the manifest file that are useful for running applications. The compiled source code from each application is saved as the \textit{classes.dex} file. 
Android application corporate machine learning (ML) algorithms to analyze the manifest information and user profiles/histories to customize the functionality and speed up the user demands~\cite{barrera2010methodology,reaves2016droid}. Also, ML algorithms utilize an Android application to detect anomalies and malware software~\cite{peng2012using}. The aim of the malware as a malicious software in mobile applications is to steal confidential data and to obtain root privileges of each APK~\cite{roy2015experimental}. Malware authors (i.e., adversaries) look for the length of malware propagation cycle to launch attacks on ML-based detectors~\cite{moser2007limits}. To accomplish this, malware applications are repackaged with attributes extracted from legitimate programs to evade detection~\cite{huang2011adversarial,lindorfer2015marvin,zhou2019deep}. In a nutshell, the generated malware sample is statistically identical to a benign sample. 
To do so, adversaries adopt \textit{adversarial machine learning algorithms} (AML) to design an example set called \textit{poison data} which is used to fool machine learning models. Adversaries adopt several AML methods like DroidAPIMiner~\cite{aafer2013droidapiminer}, Mystique~\cite{meng2016mystique}, PIndroid~\cite{idrees2017pindroid}, and DroidChameleon~\cite{rastogi2014catch} to reduce the detection rate of classification algorithms. 

According to~\cite{papernot2016limitations}, the methods used by AML focus on two general axes: i) \textbf{Attack Complexity}: this involves reducing the complexity to craft attacks, and ii) \textbf{Attacker's Knowledge}: this is related to knowledge about architecture, training examples and algorithm to gain knowledge about the detector. If an adversary is aware of the architecture, training data or features derived from applications, the attack is called a \textit{white-box attack} (see some approaches like~\cite{grosse2017adversarial}). On the other hand, if the Adversary's knowledge is limited, then the attack is a \textit{black-box attack} (see some approaches like~\cite{goodfellow2014generative}). Adversarial classification can be \textit{False positive} or \textit{False negative}. In the former, an attacker generates a negative sample to misclassify as a positive one. On the contrary, in the latter case malware is injected with part of the benign data to bypass detection. Adversarial specificity can be \textit{targeted} or \textit{non-targeted}.

In targeted malware detection systems, an adversary can fool a classifier and predict all adversarial samples as the same class. This also maximizes the probability of a targeted adversarial class. Conversely, non-targeted adversaries can arbitrarily target a class. To do so, this group of adversaries conducts several targeted attacks and takes the one with the smallest perturbation from the results or minimizes the probability of the correct class. Finally, adversarial attack frequencies can be \textit{One-time} or \textit{multiple times/iterative}. If a set of poison data is required to be generated in real-time, adversaries should choose a one-time attack; otherwise, the attack strategy can be iterative to update the poison data. Moreover, it requires more interactions with the victim classifier, and it costs more computational time to generate them. To cope with these attacks, we need some adversarial training which injects poison data into training data to increase robustness and detect the malware~\cite{szegedy2013intriguing}.  


\subsection{Contribution}        
Different questions arise about this context, such as: How to find a way to produce poison data that will be added to the current ML model and will be unrecognized by the current anti-malware solutions? How can we leverage machine learning to improve system security by presenting some adversary-aware approaches? Do we require retraining of the current ML model to design adversary-aware learning algorithms? How to properly test and validate the countermeasure solutions in a real-world network? The purpose of this paper is to clarify these issues. To be precise, the main contribution of this research is proposing a white-box AML mechanism against \textit{poison attacks}.

To sum up, we make the following contributions:

        \begin{itemize}[leftmargin=*]
        \item We propose five different attack scenarios to generate poisoned malicious apps to disguise the learned model. 
        \item We implement two countermeasure methods as defense mechanisms that improve the detection accuracy of the compromised classifier.
        
        \item We evaluate attack and defense using three benchmark malware dataset. Additionally, we conducted theoretical analysis by estimating space and time complexity to prove the scalability of our approach.
        
        \item Furthermore, we compared the proposed attack scenarios against the state-of-the-art method employing the Jacobian matrix~\cite{grosse2017adversarial}. Moreover, we conclude that the attacks modeled by us have the net effect identical to prior research work, in terms of misclassification rate. 
        \end{itemize}

 \subsection{Roadmap}
    The paper is divided and arranged as follows: In Section~\ref{relatedWork}, discusses the literature review of related works. Section \ref{problemDefinition} details the preliminary different attack scenarios proposed in AML architecture and the related components. Section~\ref{PoisoningAttackDefinition} reports the proposed approaches for malware detection systems, including poison attack scenarios using AML, while Section~\ref{Countermeasure} presents the defensive strategy against attacks. Next, in Section~\ref{resultAnalysis} we present the performance analysis of proposed methods. In Section~\ref{discussion} we describe the achievement of the experiment and provide some open discussion regarding our method. Finally, Section~\ref{conclusion} concludes the paper and presents future directions of work.

\section{Related Work}\label{relatedWork}
    We divide related works into three different classes: i) AML methods in different contexts which we present them in Section \ref{sec:2.1}, ii) AML in Android malware that we exemplify them in Section \ref{sec:2.2}, and iii) AML applies in Android malware with static features and their presented countermeasures that we add them in Section \ref{sec:2.3}. Then, we describe the works that fall into each class.

\subsection{AML in general}\label{sec:2.1}

Adversaries apply complex algorithms to generate small perturbations on original datasets in order to increase the probability of fooling ML algorithms. Some of the most efficient and important methods are presented in \cite{szegedy2013intriguing,carlini2018audio,shen2016uror,goebel2018explainable,kurakin2016adversarial,eykholt2018physical,kreuk2018fooling}. For the first time, authors in \cite{szegedy2013intriguing}, introduce a set called \textit{adversarial examples}\footnote{In this paper, poison data is used interchangeability as the adversarial example.} that have directly emerged as an input to a classification model using gradient-based optimization. This set of data is similar to misclassified data and shows that they can target a class without emulating any of the human-perceptible attributes of that class. Recently, several authors have incorporated adversarial examples for different realistic case studies, such as the audio adversarial example reported in~\cite{carlini2018audio}, manipulating a traffic sign recognition system to detour autonomous vehicles~\cite{kurakin2016adversarial}, and a perturbed physical objects model to evade object detection methods~\cite{eykholt2018physical}. Recently, Papernot et al.~\cite{papernot2014} study the positive effect of attack distillation for neural networks (NN) and propose a defense mechanism against adversarial perturbations.

\vspace{-5px}
\subsection{AML in Android malware}\label{sec:2.2} 

Having applications initially requesting less permission during installation time can defeat a machine-learning system based on permissions. In particular, an adversary may create malicious applications to have a uniform distribution of permissions as in a benign dataset. Since static analysis necessarily does not extract all the Android apps features, the training process of the detection model will not be trained based on all of the features.  Therefore, in cases where malware has been distinguished from benign by these additional features, the proposed models do not have the ability to discriminate between them. Consequently, the developed models will yield a higher misclassification rate. Studies in~\cite{Moonsamy2012} report the extraction of sensitive data from devices with apps demanding zero permission during installation.
Meanwhile, the authors in~\cite{Narain2016} illustrate that a zero permission app could be used to infer the user’s location and routes traveled using an accelerator, magnetometer, and gyroscope. Besides, the authors in \cite{grosse2017statistical} present a new hypothesis which identifies adversarial inputs based on classifier output. They validate this method using statistical tests before they are even fed to the ML model as inputs. The approach is exciting and moves one step further on adversarial example appliances in ML. However, they did not present any countermeasure to manage such malicious behavior. Conversely, we introduce two re-training defense strategies to mitigate this limitation. Moreover, in~\cite{biggio2013evasion}, the authors present a classification system which helps adversaries to craft misclassified inputs and easily evade a deployed system. This attack method, which is established during the test phase, learns to increase the attacker’s knowledge of the system, is classified as a targeted iterative attack, and helps the attacker to manipulate attack samples. This is a skillful method that increases the attacker's flexibility and performance. It also injects poison examples into training data to fool the learning algorithm and causes misclassification errors. However, it was only tested on one dataset and did not discuss countermeasure for such a white-box attack, while our paper addresses these aspects.

\vspace{-15px}
\subsection{AML in Android countermeasures}\label{sec:2.3}

The authors in~\cite{goodfellow2014generative} present two new adversarial models inspired by \textit{Generative Adversarial Network (GAN)}~\cite{goodfellow2014generative} that is based on minimax two-player between the generator (adversary) and the discriminator (classifier or system). In other words, GAN is a game which terminates at a saddle point that is a minimum with respect to the generator and a maximum with respect to the discriminator.

 In another work~\cite{grosse2017adversarial}, the authors adopt adversarial examples to construct an effective attack against malware detection models. Unlike the previous solution, such attack directs discrete and binary input datasets, like the DREBIN dataset, which is a targeted, iterative white-box method. The interesting point in this method is that it also incorporates some potential defense mechanisms, e.g., defensive distillation~\cite{papernot2016distillation} and adversarial training~\cite{kurakin2016adversarial}, using deep neural networks (DNNs) to handle malware detection models. Their achievements indicate that their countermeasures provide robustness based on the perturbation of the distribution and reduce the rate of misclassified adversarial examples. The paper is of some interest; however, our method has two significant benefits compared with this method. First, our method uses a different type of attack strategies on various datasets, which enables adversaries to easily target the discrete domains based on the form of the distributed datasets. Second, we use GAN and retraining as defense strategies to improve detection rate.

Recently, in ~\cite{Huang2019}, the authors analyze white-box and grey-box attacks to an ML-based malware detector and conduct performance evaluations in a real-world setting. Their main goal is to investigate the vulnerabilities of an ML-based malware detector and generate some countermeasures for such type attacks. In their attack scenario, they can bypass the real-world ML-based malware detector only by modifying one bit in the feature vector.

\section{Preliminaries}\label{problemDefinition}
In the following, we briefly introduce data modeling for static malware detection methods in Section~\ref{sec:3.1} and ML-based detection methods in Section~\ref{sec:3.2}. 

\subsection{Data modeling for static malware detection}\label{sec:3.1}

The purpose of constructing adversarial instances in malware detection systems is to fool the classification algorithms used by these systems and cause that system work in the way the attacker intends. In this paper, we consider the standard setting for designing a classifier in a problem that includes discrimination between benign ($B$) and malware ($M$) samples. In this way, we first select the learning algorithms and performance evaluation settings. Then, we collect a dataset $D$ that includes $n$ labeled examples and extract $m$ features for each sample. Hence we have

\begin{equation}
\label{eq:eq1}
D=\{(x_i,y_i)\mid \forall i=1,\ldots,n\},
\end{equation}
where $x_i$ is the $i$th sample vector of a dataset in which each element shows the selected feature and $y_i$ is the related label of the samples, and $y_i \in \{0, 1\}$. Let $x_{ij}$ be the value of the $j$th feature in $i$th sample where $\{\forall j=1,\ldots,m\}$. If vector $x_i$ has the $j$th feature then $x_{ij} = 1$; else $x_{ij} = 0$. In this definition $n$ is the number of samples in dataset, and $X\subseteq \{0,1\}^m$ is a feature space with $m$ dimension.

Furthermore, we consider binary classification algorithms in which the adversary changes the malicious dataset to prevent detection. The adversary tries to change the malicious data set $ y_i = 1 $ in each direction by adding a non-zero value to the feature vector. For instance, adversaries may add benign-related features to the only malware samples to evade detection by classifiers. Therefore, we can construct an adversarial example $x^*$ from a benign sample $x$ which is misclassified by the classifier $F$ and present it in equation~\eqref{eq:eq02}:
\begin{equation}
\label{eq:eq02}
X^*=X+\delta_x, \quad \text{s.t.} \quad F(X+\delta_x)\neq F(X),
\end{equation}
where $\delta_x$ is the minimum value can used as a perturbation and cause misclassification.
\subsection{Machine learning based detection method}\label{sec:3.2}
Goodfellow et al. ~\cite{Goodfellow2014} validate that practical attack in deep neural networks is possible because these models are locally linear. Also, they confirm that boosted trees are even more susceptible to attack than neural networks. Therefore, it can be a good reason that we apply our attack scenarios on existing tree type classifiers. Hence, in this paper, we use three classifiers: Random Forest, Bagging, and SVM. We provide a short description of each of them in the following: \begin{itemize}[leftmargin=*]
    \item {\textbf{Random Forest (RF).} } RF is a machine learning algorithm that creates multiple decision trees and combines the results to provide more accurate and reliable predictions. 
    \item {\textbf{Bagging.} } A Bagging classifier is an estimator that combines the base classifier results on random sets and builds an ensemble learning-based classifier. Each classification training set is randomly generated, with replacement. 
    \item {\textbf{Support Vector Machine (SVM).} } Support Vector Networks are learning models with supervised learning algorithms that inspect the data used for classification and regression analysis. 
\end{itemize}

\section{Adversarial Approaches for Malware Detection System}\label{PoisoningAttackDefinition}

 We define five different scenarios which are detailed in the following subsections. Our attack scenarios are a targeted attack. This means that the attacker generates some misclassified samples to infect a particular device. The main notations and symbols used in this paper are listed in Table~\ref{tab1}.

\begin{table}[!htpb]
\centering
\caption{\small Notation and symbols used in this paper.\vspace{-5px}}
\label{tab1}
\scriptsize{
\setlength\tabcolsep{1.5pt} 
\begin{tabular}{|c|
>{\columncolor[HTML]{fcfcf4}}l|}
\hline
\rowcolor{LightCyan}
\textbf{Notations}  
&\multicolumn{1}{|c|}{\textbf{Description}}\\ \hline
\cellcolor[HTML]{E8E8AB}\textbf{$X$}& Input data (unmodified data), $x\in \{0,1\}^m$, $\mid X \mid=n$ \\\hline
\cellcolor[HTML]{E8E8AB}\textbf{$Y$}& Label of class in the classification problem, $y\in \{0,1\}^m$ \\\hline
\cellcolor[HTML]{E8E8AB}\textbf{$X^*$}& Adversarial example (modified input data)\\\hline
\cellcolor[HTML]{E8E8AB}\textbf{$Y^*$}& Label of adversarial class in target adversarial example \\\hline
\cellcolor[HTML]{E8E8AB}\textbf{$M$}& Number of features, $a\in M$, $\mid M \mid=m$ \\\hline
\cellcolor[HTML]{E8E8AB}\textbf{$f$}& ML model, $f:X\rightarrow Y$ \\\hline
\cellcolor[HTML]{E8E8AB}\textbf{$\theta$}& Parameters of ML model $f$ \\\hline
\cellcolor[HTML]{E8E8AB}\textbf{$\lambda$}& Percentage of features\\\hline
\end{tabular}}
\end{table}

\vspace{-15px}
\subsection{Attack strategy and scenarios}

The attack strategy defines how the attacker compromises the system, based on the hypothesized goal, knowledge, and capabilities. In this paper, we characterize the attacker's knowledge in terms of a set $S$ that encodes knowledge of the data $X$, the feature set $M$ and the classification function $f$. Furthermore, we assume the attacker has complete knowledge of the target system, and formally represent the model as $V^*$ = $( X;M; f)\in S$. We assume that the initial set of input data (i.e., or samples) $X$ is given. The attack strategy is to modify the data samples using a modification function $\Omega(X)$. 

Assume that the attacker's knowledge $x^*$, we define a set of manipulated attacks as $X^*\in\Omega(X)\subseteq \mathcal{Z}$, then, we can define the attacker's goal as an objective function $W(X^*, V^*)\in \mathcal{R}$ which evaluates the extent to which the manipulated attacks $X^*$ meet the attacker's goal. Hence, we can define the optimal attack strategy as:
\begin{equation}
\label{eq:eq2}
OPT(X^*)=\text{arg\ max}_{X^*\in \Omega(X)}\: W(X^*, V^*).     
\end{equation}

To this end, we summarize each attack scenario as follows:

\noindent\textbf{Scenario 1:} The attacker randomly manipulates the features of the malware applications without having knowledge of which feature is prominent; we call this a \textit{trivial attack}.

\noindent\textbf{Scenario 2:} The attacker manipulates the malware instances in training set by altering features statistically relevant in the legitimate application distribution; we call this a \textit{Distribution attack}.  

\noindent\textbf{Scenario 3:} The attacker computes the similarity of the malicious sample with the distribution of benign samples and tries to modify those samples which are closest to the $k$ (e.g., $k=10$) nearest benign samples; we called this $KNN$ \textit{based attack}.

\noindent\textbf{Scenario 4:} The attacker manipulates the feature vector corresponding to each sample using the logistic regression function (LR) that fits the data points. We select those data points which are close to the benign feature vector. Such well-crafted attacks are referred to us as \textit{Logistic Regression attack} or $LR$. 

\noindent\textbf{Scenario 5:} The attacker manipulates the malware instances by adopting the LR function and bio-inspired solution to find a global solution. In this scenario, we adopt ant colony optimization (ACO) as a sample of the bio-inspired method to produce poison malware samples found close to goodware. We name this attack as \textit{ACO attack}. 

We repeat each algorithm ten times and select the average values for each parameter. For all these scenarios, we divide the training and test datasets based on the class parameters into  \textit{Malware} and \textit{Benign} datasets. 
Subsequently, we apply feature ranking on benign examples and select 10\% of the top-ranked features. The ranking reflects those attributes which can classify an unseen sample to benign class with high probability. Furthermore, we choose some percentages of the selected features among the selected malware samples and modify the feature values. Formally speaking, we select the feature with zero value which has not been selected before and changes zero to one. At this stage, we add such modified malware samples to the test dataset and classify the dataset using classification methods.

\subsection{{Trivial attack}}\label{attack1}

In Alg.~\ref{alg:algorithmTrivial}, focusing on line 1 and 9-18, the list variable $\lambda$ explains the features applied. To be precise, first, we select three features from the list variables and modify them; then, we repeat this process on all of the members in the feature list $\lambda$. In the loop, for each selected sample, we check the selected feature $a$ in the $\lambda$ set and change the zero values of the feature to one and save that sample in $x^*$ (line 8-10 of Alg.~\ref{alg:algorithmTrivial}). Then, it is important to check the modified sample and understand if it changes to a benign sample or not. We use  $F(x^*)=y^*$ to check this condition.  If it is satisfied, we can call $x^*$ as an adversarial sample (i.e., poison sample) which is the output of  Alg.~\ref{alg:algorithmTrivial}.

\begin{algorithm}[t]
\caption{\small Scenarios 1-4:  Trivial, Distribution, KNN, and Logistic Regression attacks.}
\label{alg:algorithmTrivial}
\textbf{Input:} $x$, $y$, $\lambda$\\
\textbf{Output:} 
 $x^*$, $y^*$
\begin{algorithmic}[1]
\color{black}
\footnotesize
\Statex{\texttt{Scenario 1: Trivial Attack.}}
\State{Randomly select $\lambda$ features from $X$}
\Statex{\texttt{Scenario 2: Distribution Attack.}}
\State{Randomly select ranked($\lambda$) features from $X_b$}
\Statex{\texttt{Scenario 3: KNN Attack.}}
\State{Ranked($\lambda$) features from $X_b$}
\State{$X_s\leftarrow$ Randomly select $10\%$ of samples from $X_m$}
\For{$\forall x\in X_s$}
\State{$x \leftarrow$ Find KNN of samples from $X_b$}
\EndFor
\Statex{\texttt{Scenario 4: Logistic Regression Attack.}}
\State{$discriminator\leftarrow$Randomly select $\lambda$ features from $X_b$ using \textit{LR}}
\State{$x\leftarrow$select $10\%$ of $X_m$ near $discriminator$}
\Statex{\texttt{Common parts for Scenario 1-4.}}
\For{each attribute $a$ $\in$ $\lambda$ in $x$}
	\If{($x[a]=0$)}
	   \State{$x[a]\leftarrow 1$}
     \State{$x^* \leftarrow x$}
     \If{($F(x^*)=y^*$)}
        \State{break}
    \EndIf
     \EndIf
\EndFor
\State{\textbf{return} $x^*$, $y^*$}
\end{algorithmic}
\end{algorithm}

\subsection{{Distribution attack}}\label{attack2}

In this scenario, we randomly manipulate the selected ranked features of malware samples placed in the malware dataset (see line 2 and loop lines 9-18 of Alg.~\ref{alg:algorithmTrivial}). After the modifications, we feed the modified malware with the benign sample of the test dataset to the learning model and update the learning parameters. 

\subsection{{KNN based attack}}\label{attack3}

In this scenario, we first rank the $\lambda$ features of the samples in the benign dataset. Second, we randomly select 10\% of the malware data (samples). We calculate the Manhattan distance of each selected sample with benign files. Moreover, we select the $K$ (in this paper, we consider $k=10$) nearest benign vectors to the corresponding sample the $\lambda$ highest ranked features. Indeed, we have $k$ new poison samples for each malware sample. We should recall that the ranking is used to understand the highest features in the benign samples. Then, we add these poison samples with benign data from $x$ to the learning model (see line 3-7 and loop lines 9-18 of Alg.~\ref{alg:algorithmTrivial}).

\subsection{{Logistic regression attack}}\label{attack4}
In this scenario, we apply an LR algorithm on the training dataset, and the result of this algorithm will be the \textit{discriminator}. The nearest malware samples to this discriminator are the best choices for the modifications. Therefore, we select 10\% of the malware samples in the training dataset which are near to the discriminator. After that, we compare the selected malware (selected based on 10\% of the malware samples in the training dataset) with the malware samples in the test dataset and add ten malware samples to the test dataset for each malware sample in the training dataset. In this step, for 10\% of the training dataset, we select the samples with all zero features and apply logical `OR' with randomly selected malware samples in the same dataset. The resulting sample will be poison data which can be used for the classification (see line 8-7 and loop lines 9-18 of Alg.~\ref{alg:algorithmTrivial}). 

\subsection{{Ant colony based attack}}\label{attack5}

In this scenario, new adversarial samples are generated using an ant colony optimization (ACO) algorithm (see line 1 of Alg.~\ref{alg:algorithmACO}). First, we apply a linear regression algorithm to select the malware samples which are most similar to the benign samples in the training dataset (i.e., 10\% of the malware samples in the training dataset). Formally speaking, we find the nearest malware to the discriminator in the training dataset (i.e., we search only in the 10\% of the malware samples in the training dataset) (see line 5 of Alg.~\ref{alg:algorithmACO}). In Alg.~\ref{alg:algorithmACO}, the function $ACOFunction$ is used to find the adversary sample data. In this way, the ACO pheromone value is the number of the feature that is going to be changed. First, we start the ACO with one feature, and we generate new samples by modifying the malware samples with the absence of attributes which are present in legitimate applications. We repeat this action by using more features. If the distance between the newly generated sample and the discriminator is within the range of the selected malware and the discriminator, then we add this newly generated sample to the recently generated samples; otherwise, we discard this sample and change the feature and re-calculate the distances. We continue this process until the maximum iteration is reached or the classifier misclassifies malware samples. The algorithm of the ACO scenario attack is described in Alg.~\ref{alg:algorithmACO}.
\begin{algorithm}[t]
\caption{\small ACO scenario}
\label{alg:algorithmACO}
\textbf{Input:} $X$, $Y$, $\lambda$, $ACOthreshold$\\
\textbf{Output:} 
 $X^*$, $Y^*$
\begin{algorithmic}[1]
\color{black}
\footnotesize
\State{$discriminator\leftarrow$Randomly select $\lambda$ features from $X_b$ using \textit{LR}}
\State{$X_s\leftarrow$select $10\%$ of $X_m$ near $discriminator$}
\For{each $x$ $\in$ $X_s$}
\State{$x^*\leftarrow ACOFunction(x)$}
	\If{($distance(x^*,discriminator)\leq ACOthreshold$)}
	 \If{($F(x^*)=y^*$)}
    \State{$X^*\leftarrow X^*\cup x^*$}
    \EndIf
        \EndIf
\EndFor
\State{\textbf{return} $X^*$, $Y^*$}
\end{algorithmic}
\end{algorithm}

\section{Defensive Strategies Against Attacks}
\label{Countermeasure}
In this section, we discuss two countermeasures as the main solution for the raised attacks.

\subsection{{Adversarial training}}
\label{RetrainingCountermeasure}

\begin{figure}[!htpb]
    	\begin{center}
	\includegraphics[width=\columnwidth]{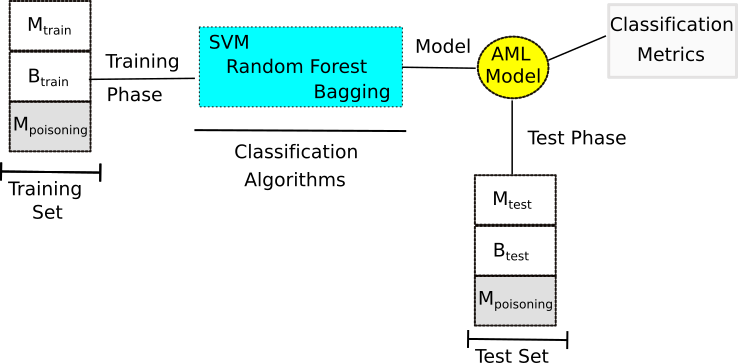}
    		\caption{\small Adversarial training model.}
    		\label{fig:fig2}
    	\end{center}
       \end{figure}

In the first defense method, we re-train the classification algorithm~\cite{kurakin2016adversarial}. The main difference between the new re-trained classification dataset and the current version is that we add the poison data with the training dataset. Figure~\ref{fig:fig2} presents the structure of the adversarial training countermeasure. In Fig.~\ref{fig:fig2}, the left-side boxes illustrate the training set which is used as a training model and the lower-side boxes are used as a testing phase for the learning model. The presented model uses different classification algorithms such as SVM, Bagging, and Random Forest in this paper. 

\begin{algorithm}[t]
\caption{\small Adversarial training: pseudo-Code of the retraining defense}
\label{alg:algorithmDef1}
\textbf{Input:} $X$, $Y$, $X^*$, $Y^*$ \\
\textbf{Output:} $Model_{new}$
\begin{algorithmic}[1]
\footnotesize
\State{$X_{trn},\ Y_{trn}\leftarrow$ Randomly select 60\% data from $X$ and $Y$}
\State{$X^*_{trn},\ Y^*_{trn}\leftarrow$ Randomly select 60\% data from $X^*$ and $Y^*$}
\State{$X_{trn}\leftarrow X_{trn}\cup X^*_{trn}$}
\State{$Y_{trn}\leftarrow Y_{trn}\cup Y^*_{trn}$}
\State{$Model_{new}= RandomForestClassifier(X_{trn},Y_{trn})$}
 \State{\textbf{return} $Model_{new}$}
\end{algorithmic}
\end{algorithm}

We present the steps of retraining method in the Alg.~\ref{alg:algorithmDef1}. The input of this algorithm is the dataset, and corresponding labels of adversarial examples (i.e., poison data) and the original dataset, and the output will be the new retrained model using the Random Forest classifier~\cite{ho1995random}. First, we randomly select 60\% of the dataset and corresponding labels and save them in the training subset for the original and poison data (See lines 1 and 2 of Alg.~\ref{alg:algorithmDef1}). Then, we build the new training data as presented in lines 3 and 4 of Alg.~\ref{alg:algorithmDef1}. Next, we feed the adversarial model using such new trained data with the help of the \textit{RandomForest Classifier} and retrain the model. The new model is used to implement data classification. It is rational that the new model which is produced by the poison data and the preliminary dataset has higher precision in data classification compared to the existing model. 

\subsection{{Generative adversarial network (GAN)}}
\label{GANCountermeasure}

\begin{figure}[!htpb]
    	\begin{center}
    		\includegraphics[width=\columnwidth]{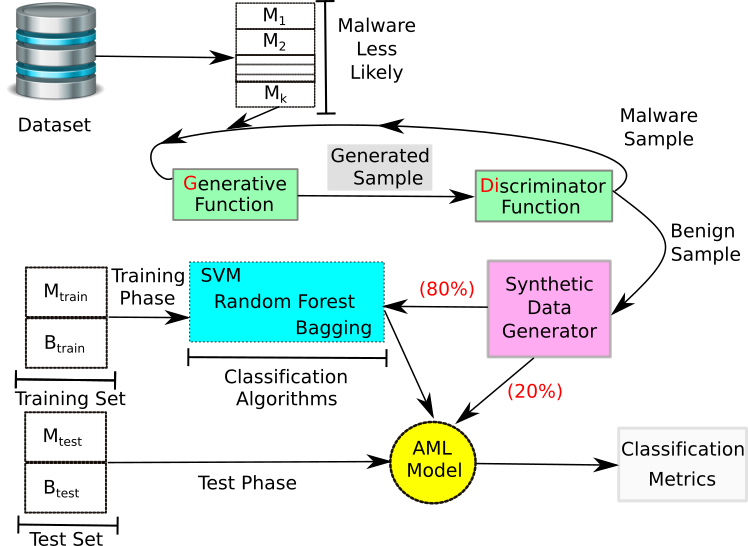}
    		\caption{GAN defense architecture.}
    		\label{fig:fig3}
    	\end{center}
       \end{figure}

In the second defense method, which is called \textit{generative adversarial network} (GAN), we exploit the Random Forest Regression to select 10\% of malware samples that have the greatest similarity to the benign samples in the training dataset and generate a \textit{ less likely} malware set. We use the GAN as a synthetic data generator set. The GAN has two functions called \textit{Generator} and \textit{Discriminator}. The generator function is used to modify the less likely malware samples. To do so, one random feature from the highest ranked features with zero value in the training dataset is selected and its value is changed to one and generates a new sample. The new sample is fed to the second function, the discriminator -- which works like a classifier -- to predict the class variable. The discriminator module modifies the features until the discriminator function is cheated and labels such a sample among the benign samples. Further, we gather the wrongly estimated malware samples into a synthetic data generator set. Besides, we use 80\% of the synthetic data generator set with the training dataset to update the adversarial learning model. We use the remaining synthetic data generator samples (i.e., 20\% of the data samples) with the test dataset to analyze the classification. Figure~\ref{fig:fig3} presents the GAN defense architecture. \footnote{GAN is also can be used to generate adversarial example and fool the classifier which is out of the scope of this paper.}     
       
\begin{algorithm}[t]
\caption{\small GAN: pseudo-code of the GAN defense}
\label{alg:algorithmDef2}
\textbf{Input:} $X$, $Y$, $X^*$, $Y^*$, $\lambda$ \\
\textbf{Output:} $Model_{new}$
\begin{algorithmic}[1]
\footnotesize
\State{$Model_{poison}\leftarrow$ Fit a model on $X_{trn}$ using $LR$}
\State{$Lesslikely\leftarrow$ Fit 10\% model of $X_{m}$ with KNN to $Model_{poison}$}
\For{each $x\in Lesslikely$ }
\State{$x_{new}\leftarrow x$}

\While{$x_{new} \in Model_{poison}$ Classify as $M$}
\State{$x_{new}\leftarrow$ Add ranked{($\lambda$)} features from $X_{b}\cup\ x_{new}$}
\EndWhile
\State{$synthetic_{data}\leftarrow synthetic_{data}\cup x_{new}$}

\EndFor
\State{$Model_{new}\leftarrow$ Fit Model on $X_{trn}\cup synthetic_{data}$}
 \State{\textbf{return} $Model_{new}$}
\end{algorithmic}
\end{algorithm}   

It uses a generative adversarial network (GAN) for building new samples. The GAN applies two neural networks which use the back-propagation technique. One network generates candidates (called \textit{generative}), and the other evaluates them (called \textit{discriminator}). In the GAN, the training process is applied in the generative function in such a way as to increase the error rate of the discriminator function. The GAN structure is a competitive setting, i.e., versus the discriminator function and simplified Turing learning~\cite{li2016turing}.

Alg.~\ref{alg:algorithmDef2} clearly explains the steps of the GAN-wise countermeasure. In detail, we give the samples as an input set to the generator function. We modified these samples using the important features of benign samples to fool the discriminator function by producing novel synthesized instances that appear to have come from the benign dataset. In other words, the discriminator function’s task is to discover the benign samples from the malware ones. If the discriminator function correctly recognizes a malware sample, that sample returns to the generator function to re-train the sample. In this algorithm, we first generate a poison model, called \textit{Model_{poison}}, using the logistic regression algorithm (see line 1). Then, in line 2 of Alg.~\ref{alg:algorithmDef2}, we build 10\% samples of the malware dataset which are KNN nearest to the established poison model and save these samples in \textit{Lesslikely} set. Indeed, \textit{Lesslikely} set is composed of samples that have the greatest similarity to the malware dataset. For each sample in \textit{Lesslikely}, if it belongs to the poison model we label it as `M' for Malware, then we try to modify it with the most important features in the benign dataset until it falls out from the malware dataset and is classified as part of the benign dataset (see the lines 5-7 of Alg.~\ref{alg:algorithmDef2}).

Hence, if a sample in GAN is recognized as a benign sample, we add it to the \textit{synthetic\_data} set. After analyzing all the samples, we use 80\% of the \textit{synthetic\_data} samples set with the training dataset (malware and benign) to generate the new model and feed it to the classification algorithms. The remaining samples are in the \textit{synthetic\_data} with a test dataset for analyzing the new model. In a nutshell, we will see that the new model improves the level of accuracy.

\subsection{Time/Space complexity of the attack and defense algorithms}\label{ComplexityAttacksDefenses}

In the following section, we conduct time and space complexity analysis on the presented attacks and defenses. 

\subsubsection{Time complexity}\label{TimeComplexityAttacksDefenses}

Focusing on the time complexity of Alg.~\ref{alg:algorithmTrivial}, we consider four attack scenarios which have similar code interactions from lines 10--19. Note that this common part (\textit{for} loop presents in lines-10--19) runs off the order of $\mathcal{O}(\lambda\cdot n)$ for all samples (i.e., we have a total of $n$ samples), where $\lambda$ is the selected features of our $m$ total number of the featured applied per sample in this paper. Also, we know that in malware dataset $\lambda<m<<n$. So, we can list the time complexity of each attack scenario as follows:
\begin{itemize}[leftmargin=*]
  \item \textbf{Trivial attack.} In this scenario, we select $\lambda$ features from a total of $m$ features in the original dataset (i.e., $X$). In the worse case, it runs off the order of $\mathcal{O}(\lambda\cdot n)$ for all samples. As a result, the overall complexity of a trivial attack is about $\mathcal{O}(m\cdot n)$.

    \item \textbf{Distribution attack.} In this scenario, we put the feature in descending order and select the $\lambda$ highest ranked features. Hence, in the worse case, it runs off the order of $\mathcal{O}(m^2)$ per sample. As a result, the overall complexity of distribution attack, considering the common block of the Alg.~\ref{alg:algorithmTrivial}, is about $\mathcal{O}(m^2\cdot n)$. 
    
    \item \textbf{KNN based attack.} In this scenario, we first select $\lambda$ features of the ordered vector of $m$ features (the time takes is of the order of $\mathcal{O}(m^2)$). Second, we randomly select 10\% of the malware samples (the time takes is of the order of $\mathcal{O}(n)$). Finally, we select the $K$ sample, so the time taken for selecting is of the order of $\mathcal{O}(k\cdot \lambda\cdot n)$ ~\cite{KNN}, where we use $k$ as a fixed value in this scenario. As a result, the overall time complexity of the KNN attack is about $\mathcal{O}(k\cdot \lambda\cdot n)$.
    
    \item\textbf{LR based attack.} In this scenario, we randomly select $\lambda$ important features from the benign dataset using the LR algorithm and select 10\% samples of the malware dataset which are near to the legitimate samples, which takes of the order of $\mathcal{O}(n)$. Considering the common block of Alg.~\ref{alg:algorithmTrivial}, the overall time is taken for the LR attack is about $\mathcal{O}(\lambda\cdot n)$. 
    
    \item\textbf{ACO attack.} In this scenario which is detailed in Alg.~\ref{alg:algorithmACO}, we first build a discriminator and select 10\% of the malware samples near the discriminator, which takes of the order of $\mathcal{O}(n)$. The main \textit{for} loop $n$ times run the ACO algorithm. Therefore, the overall time taken for the ACO attack is about $\mathcal{O}(n^4)$).
\end{itemize}

We present two defense mechanisms in this paper. In each method, we try to modify the current model and build an updated model that has more intelligence against the attacks mounted. In the following, we describe the time it takes to build the new retrained model: 

\begin{itemize}[leftmargin=*]
\item\textbf{Adversarial training defense.} In this method, which is detailed in Alg.~\ref{alg:algorithmDef1}, we first randomly select 60\% of original and poison data and build new training set which runs for around $\mathcal{O}(n)$. Then, we gather the trained and poison data which run for around $\mathcal{O}(n)$. Then, we run the Random Forest regression function on these two datasets and build a new model. The Random Forest regression function utilizes a decision tree for classifying the data and identifies the important features based on the results of the classification. Since we have $n$ samples and $m$ features, the time taken to establish each tree is about $\mathcal{O}(m\cdot n\cdot log n)$. As a result, the overall time taken for the adversarial training defense method is in the order of $\mathcal{O}(m\cdot n\cdot log n)$. 

\item \textbf{GAN defense.} According to Alg.~\ref{alg:algorithmDef2}, building the poison model takes in the order of $\mathcal{O}(n)$. Then, we use the KNN method to make a \textit{Lesslikely} matrix which takes for about $\mathcal{O}(m\cdot n)$. The loop runs for each sample out of \textit{Lesslikely} matrix and for each of them while the instruction is checked on the poison model (see lines 5-7). Hence, the duration of the \textit{for} loop runs for about $\mathcal{O}(n^2)$. As a result, the overall time taken for the GAN defense method is in the order of $\mathcal{O}(n^2)$.    

\end{itemize}

\subsubsection{Space Complexity}\label{SpaceComplexityAttacksDefenses}
Focusing on the space complexity of Alg.~\ref{alg:algorithmTrivial}, for the \textit{Trivial attack}, we need a vector space to select $\lambda$ features among $m$ available features, so the required space is in the order of $\mathcal{O}(\lambda\cdot n)$ for all samples. Regarding the \textit{Distribution attack}, we also need to select $lambda$ features out of $m$ features, so the overall space complexity of this scenario runs of the order of $\mathcal{O}(\lambda\cdot n)$ for all samples. Regarding the \textit{KNN based attack}, first, we rank the features, which is applied as in-place ordering that runs of the order of $\mathcal{O}(\lambda\cdot n)$ for all samples. Then, we need a space to save 10\% of the malware dataset, which runs off the order of $\mathcal{O}(m\cdot n)$. Finally, the overall space complexity required for this scenario runs of the order of $\mathcal{O}(m\cdot n)$. Regarding the \textit{LR attack}, LR takes at most a dataset space to randomly select $\lambda$ features in the benign dataset and occupy $\mathcal{O}(m\cdot n)$ space to find 10\% of nearest malware dataset to the discriminator. As a result, the overall space complexity of the LR scenario is in order of $\mathcal{O}(m\cdot n)$. Focusing on the space complexity of Alg.~\ref{alg:algorithmACO}, we first select $\lambda$ features out of $m$ features for each sample, so we need a space of the order of $\mathcal{O}(m\cdot n)$ for this. Then, we need to consider the order of $\mathcal{O}(m\cdot n)$ for the required space to find 10\% of the nearest malware dataset to the discriminator. In the \textit{for} loop, we build the dataset at most $n$ times. As a result, the overall space complexity of the ACO scenario is in the order of $\mathcal{O}(m\cdot n^2)$. Focusing on the space complexity of Alg.~\ref{alg:algorithmDef1} for adversarial training defense, similar to the previous algorithms, the space required for running the lines 1-4 runs for around $\mathcal{O}(n\cdot m)$. Also, the space required to generate \textit{RandomForestClassifier} function is in the order of $\mathcal{O}(NT\cdot n\cdot m)$~\cite{RandomForestComplx}, where $NT$ is the number of trees we need to consider to run this function. As a result, the overall space complexity takes about $\mathcal{O}(NT\cdot n\cdot m)$. Focusing on the space complexity of Alg.~\ref{alg:algorithmDef2} for GAN defense, similar to the previous algorithms, the space required for the \textit{LogisticRegression} function and \textit{Lesslikely} matrix is in the order of $\mathcal{O}(n\cdot m)$. All the classification algorithms using the \textit{Lesslikely} dataset can consume space in order of $\mathcal{O}(n\cdot m)$. As a result, the overall space complexity is in the order of $\mathcal{O}(n\cdot m)$. 
       
\section{Experimental Evaluation}\label{resultAnalysis}
In this section, we report an experimental evaluation of the proposed attack algorithms and their countermeasures by testing them under different scenarios. 

\subsection{Simulation setup}\label{SimulationSetup}
In the following, we present the classifiers, datasets, training/testing structure, test metrics, and hyper-\ parameter tuning.

\noindent\textbf{Classifiers.} We use three classifiers as described in Section~\ref{sec:3.2}, in which we set the k-fold variable to 10. As explained previously, the random forest (RF) algorithm classifies the data by constructing multiple decision trees. In this paper, the number of decision trees used is 100. In RF, the maximum number of features used to find the best split of the features is set to 3. In the Bagging algorithm, we use the \textit{DecisionTreeClassifier} as the base estimator. The number of decision trees used is 100. We average our results over 10 independent runs for each classifier. 
    
\noindent\textbf{Datasets:} 
We conducted our experiments using three datasets, as detailed below:
\begin{itemize}[leftmargin=*]
\item \textit{Drebin dataset:}
The Drebin dataset~\cite{arp2014drebin} is a set of Android samples that we can straightforwardly apply in a lightweight static analysis. For each Android application, we perform a linear sweep over the app's content and obtain the manifest and the disassembled dex code. We then extensively analyze all the extracted features, which are represented as a set of binary strings. These features are classified into permissions, intents and API calls. The samples contain 131,611 applications over about two years (2010–-2012) containing both benign and malware/malicious software. The Drebin dataset contains 96,150 applications from the GooglePlay store, 19,545 applications from the Chinese market, 2,810 applications from the Russian market and 13,106 samples from other sources such as Android websites, malware forums, and security blogs.

\item \textit{Genome dataset:} In Genome project which was supported by the US National Science Foundation (NSF)~\cite{jiang2012dissecting}, the authors gather about 1,200 Android malware samples from various categories from August 2010 to October 2011. They categorize them based on their installation methods, activation mechanisms, and their malicious payloads.

\item \textit{Contagio dataset:} 
Contagio mobile dataset presents a list of uploaded dropbox samples that are gathered from various mobile applications. It includes 16,800 benign and 11,960 malicious samples of mobile apps in 2015 ~\cite{Contagio}.

\end{itemize}

\noindent \textbf{Mobile application static features.} {The datasets we tested have several syntax features. The malicious applications gathered have various permissions, intents and API calls, and we assume that malicious applications are distinguishable from benign ones. We summarize the different application syntax features as follows:}
\begin{itemize}[leftmargin=*]
    \item \textit{Permission:} {Each application (APK) of an Android file has an essential profile that includes information about the application, known as permission. The Android OS needs to process these permission files before installation. This profile file indicates the permission types for each application when interacting with an Android OS or other applications. }
    \item \textit{API:} {This feature can monitor various calls to APIs in the Android OS, e.g., sending an SMS, or accessing a user's location or device ID. The Android OS provides an API framework that helps the applications to interact with the OS easily.} 
    \item \textit{Intent:} {This type of feature is used to represent communication between different components. It is also called a medium, as it can serve as a communication link between the asynchronous data exchange information and the calls to various applications.}
\end{itemize}
{In order to find the optimal number of features for modification in each attack, we repeat our experiments for $\lambda$=\{1\%, 2\%, 3\%, 4\%, 5\%, 6\%, 7\%, 8\%, 9\%, 10\%, 20\%\} of the manifest features (i.e., $M$).}
   
\noindent\textbf{Parameter setting.} {We run each attack and defense algorithm 10 times and report the average results. At each repetition, we randomly consider 60\% of the dataset as training samples, 20\% as validation samples and 20\% as testing samples. This enables us to evaluate the degree to which a classifier can maintain its detection of malware from different sources. We set the maximum iteration of the ACO attack algorithm to 1000. Besides, we modify 300 variables corresponding to the ranked feature vector using the ACO algorithm for each iteration.} We fix the evaporation ratio to 0.1 and the pheromones rate per path that is corresponding to the coefficient of experience and collective knowledge to 0.99. All of the experiments (four attacks and two defenses using three datasets with three classification algorithms) were run on an eight-core Intel Core i7 with speed 4 GHz, 16 GB RAM, OS Win10 64-bit using Python 3.6.4.
Also, we keep the source code of the paper in \footnote{\url{https://github.com/mshojafar/sourcecodes/blob/master/Taheri2019APIN-AdverserialML_Sourcecode.zip}}.

\noindent\textbf{Feature selection.} {Due to a large number of features, we first ranked the features using the \textit{RandomForestRegressor} algorithm. We then selected 300 of these features with higher ranks.}

\noindent\textbf{Comparison of solutions.} We compare our defense algorithms with the Jacobian saliency map used to craft malware samples in~\cite{grosse2017adversarial}, called \textit{JSMA}. The Jacobian-matrix-based algorithm is used to craft adversarial examples, since the binary indicator vector used by these authors to represent an application does not possess any particular structural properties or inter-dependencies. Hence, they apply a regular, feed-forward neural network with an architecture consisting of two hidden layers, each involving 200 neurons. In~\cite{grosse2017adversarial}, the authors consider at most 20 feature modifications to any of the malware applications. Our solution uses a SoftMax function for normalization of the output probabilities in the malware detection system, as follows:
\begin{equation}
\label{eq:eq3} 
F_i(X)=\frac{e^{x_i}}{e^{x_0}+e^{x_1}},\ x_i=\sum_{j=1}^{m_n}w_{j,i}\cdot x_j+b_{j,i},
\end{equation}
where $F$ is the gradient function, $x$ is the input sample, and $m_n$ is the number of features. The above authors follow two steps when building adversarial examples. In the first step, they calculate the $F$ gradient according to $X$ to estimate the direction in which the perturbation in $X$ can calculate the output of function $F$ (see \eqref{eq:eq4}): 
\begin{equation}
\label{eq:eq4} 
J_F=\frac{\partial F(X)}{\partial X}=\ \left[\frac{\partial F_i(X)}{\partial X_j}\right]_{i\in\{0,1\}, j\in[1,m]},
\end{equation}
In the second step, a perturbation $\delta$ for $X$ with a maximum positive gradient is selected in the target class $Y\textprime$. For presenting the attack mechanism, the index $i$ changes the target class to 0 by changing $X_i$, as described in \eqref{eq:eq5}:
\begin{equation}
\label{eq:eq5} 
i=\text{arg max}_{j\in[1,m], X_j=0}F_0(X_j),
\end{equation}
This process continues until one of the two following conditions is fulfilled: (i) the maximum number of changes allowed is reached, or (ii) a misclassification is successfully caused.

 \noindent\textbf{Test metrics.} 
We use certain metrics to evaluate the results, which are listed as follows:
\begin{itemize}[leftmargin=*]
\item\textit{True Positive (TP):} Denote malware correctly classified as malware.
\item\textit{True Negative (TN):} is the number of legitimate applications precisely identified by the classification algorithm.
\item\textit{False Positive (FP):} denote the number of misclassified benign applications.
\item \textit{False Negative (FN):} is the count of malware files misclassified as goodware.
\item\textit{Accuracy:} This is the ratio between the number of correct predictions and the total number of input samples. A higher value of accuracy indicates that the algorithm is able to correctly identify the label of the samples with a higher probability. Thus, we have
\begin{equation}
\label{eq:eq6} 
Accuracy=\frac{TP+TN}{TP+TN+FP+FN},
\end{equation}
\item\textit{Precision:} This is the fraction of relevant instances among the retrieved instances. Hence we have:
\begin{equation}
\label{eq:eq8} 
Precision=\frac{TP}{TP+FP},
\end{equation}
\item\textit{Recall:} This is the fraction of retrieved instances over the total amount of relevant instances. Both precision and recall are therefore based on an understanding and measure of relevance. This can be written as:
\begin{equation}
\label{eq:eq9} 
Recall=\frac{TP}{TP+FN},
\end{equation}
\item\textit{False Positive Rate (FPR):} This is defined as the ratio between the number of negative events incorrectly classified as positive (false positive) and the true number of negative events (false positive + true negative).  Therefore we have
\begin{equation}
\label{eq:eq10} 
FPR=\frac{FP}{TP+TN},
\end{equation}
\item\textit{Area Under Curve (AUC):} This defines a metric for determining the best class prediction model using all possible thresholds. Thus, AUC measures the tradeoff between (1-FPR) and FPR. The intrinsic goal of AUC is to solve the situation in which a dataset consists of unbalanced samples (or a skewed sample distribution), and it is necessary that the model is not overfitted to the class consisting of a higher number of instances. This can be written as:
 \begin{equation}
\label{eq:eq11} 
AUC=\frac{1}{2}\left(\frac{TP}{TP+FP}+\frac{TN}{TN+FP}\right).
\end{equation}
\end{itemize}
 \subsection{Experimental results}\label{results}
 
In this section, we apply the above attacks to our originally trained classifiers to examine the impact of GAN and adversarial training as defensive mechanisms against adversarial examples in the domain of malware detection.

	\begin{figure*}
    \centering
	\begin{subfigure}{0.32\textwidth}
 		\centering 
 		\includegraphics[width=\linewidth]{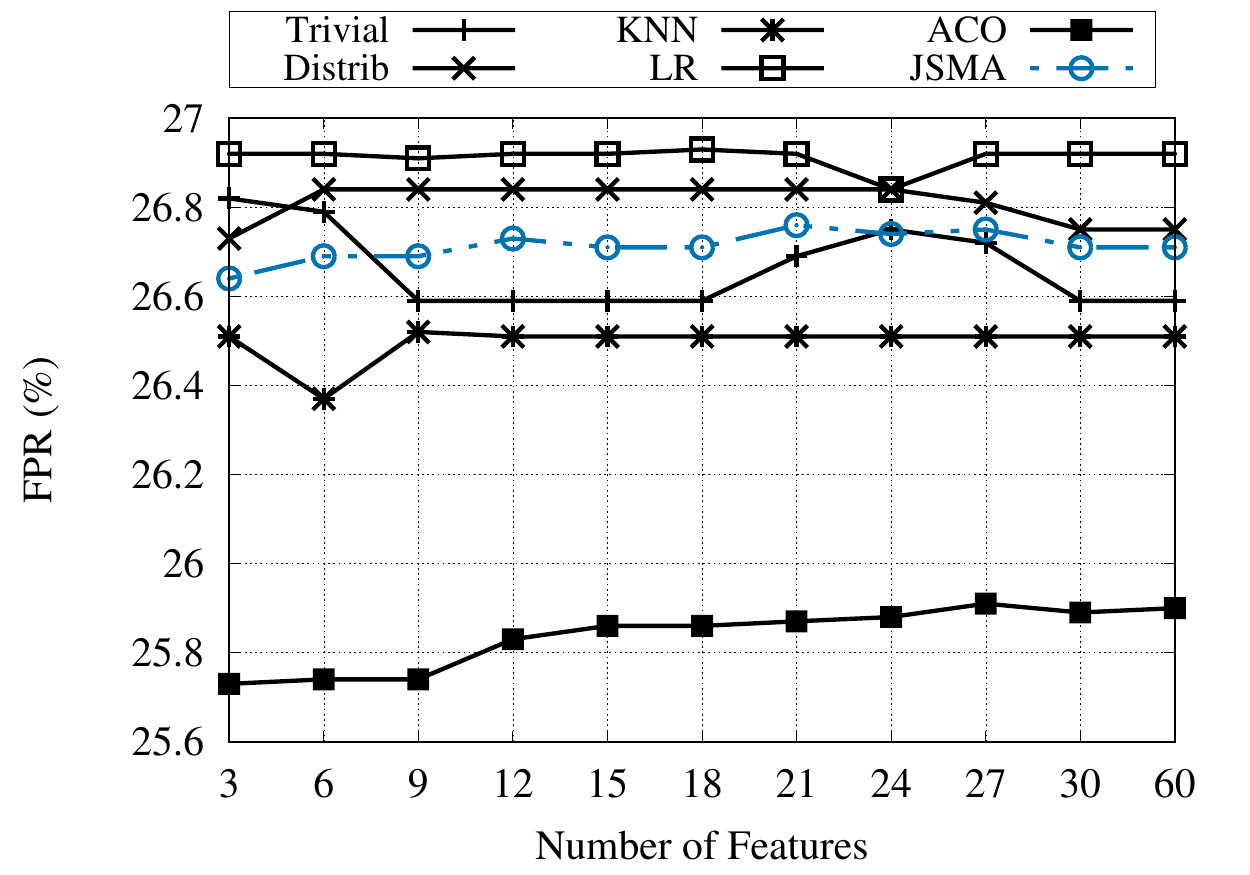}
			\caption{\small Drebin-API}
			\label{fig:fig4a}
 	\end{subfigure} 
   \begin{subfigure}{0.32\textwidth}
 		\centering
 		\includegraphics[width=\linewidth]{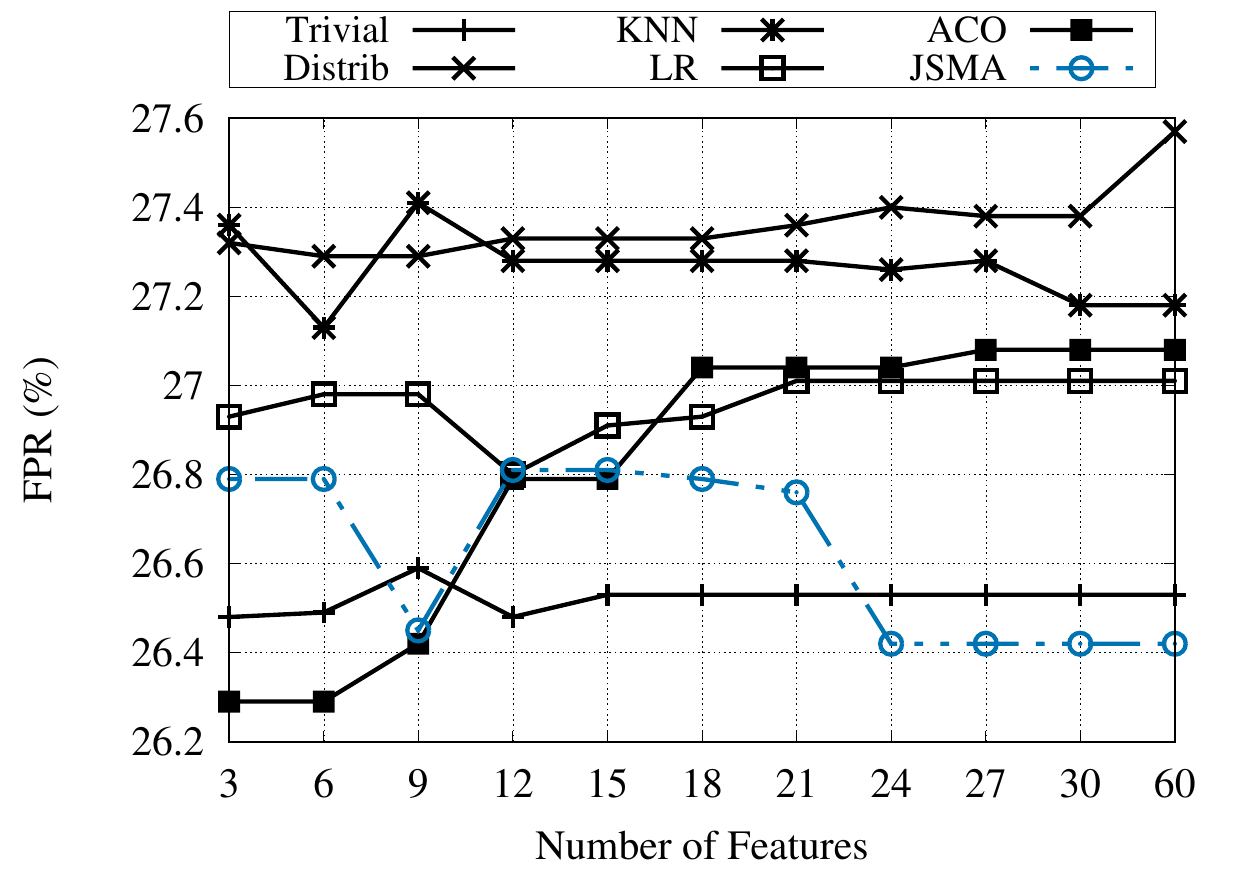}
			\caption{\small Drebin-permission}
			\label{fig:fig4b}
 	\end{subfigure} 
    \begin{subfigure}{0.32\textwidth}
 		\centering 
 		\includegraphics[width=\linewidth]{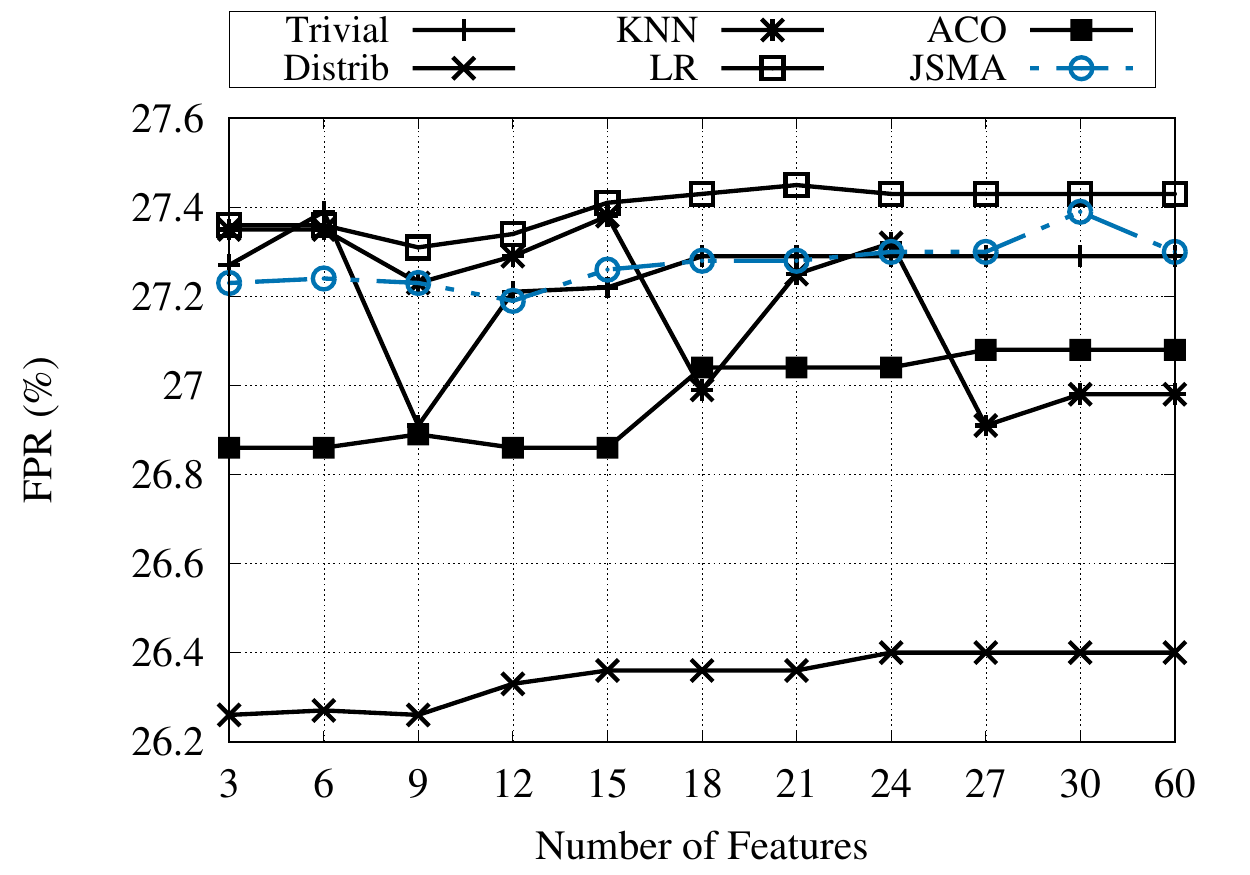}
		\caption{\small Drebin-intent}
			\label{fig:fig4c}
 	\end{subfigure}
 		\hfill
 	\begin{subfigure}{0.32\textwidth}
 		\centering
 		\includegraphics[width=\linewidth]{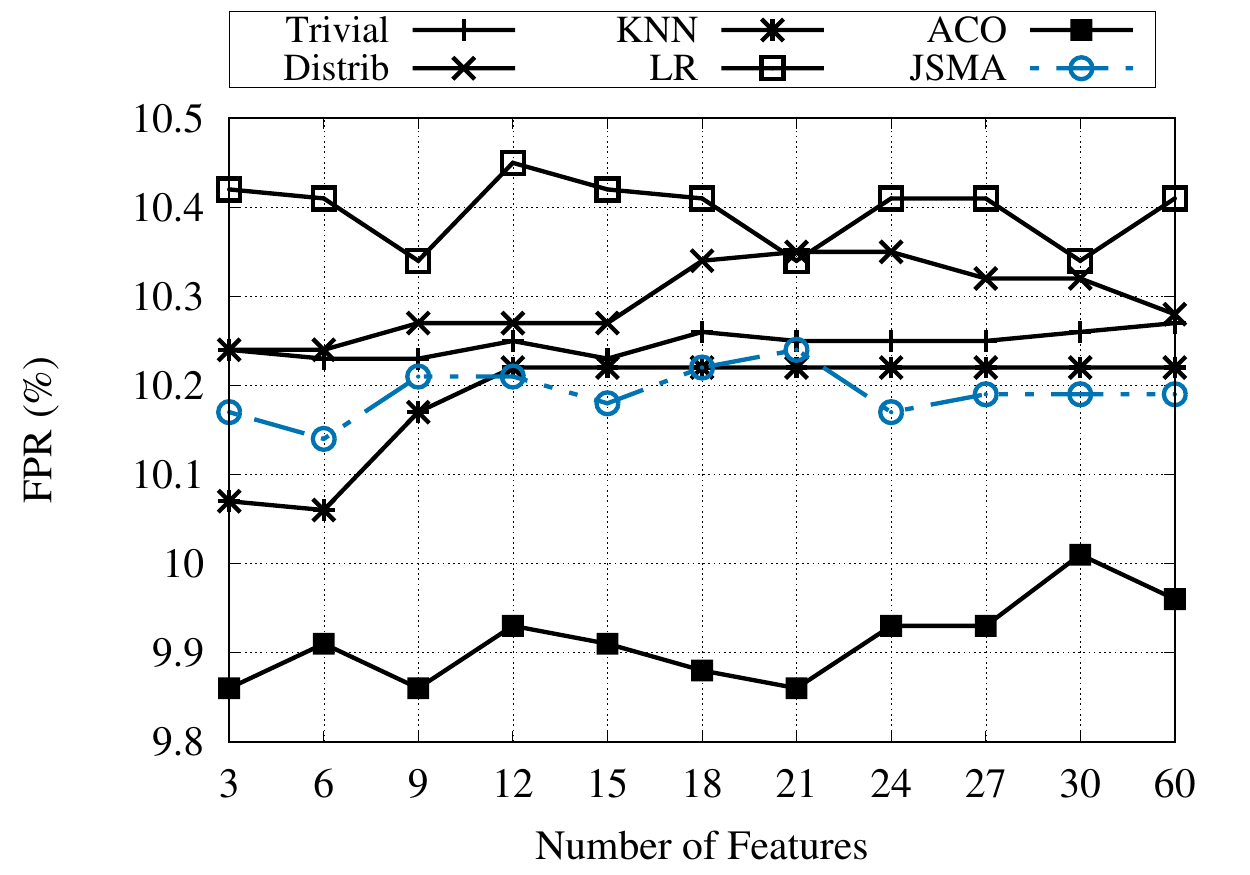}
			\caption{\small Contagio-API}
			\label{fig:fig4d}
 	\end{subfigure}  
    \begin{subfigure}{0.32\textwidth}
 		\centering 
 		\includegraphics[width=\linewidth]{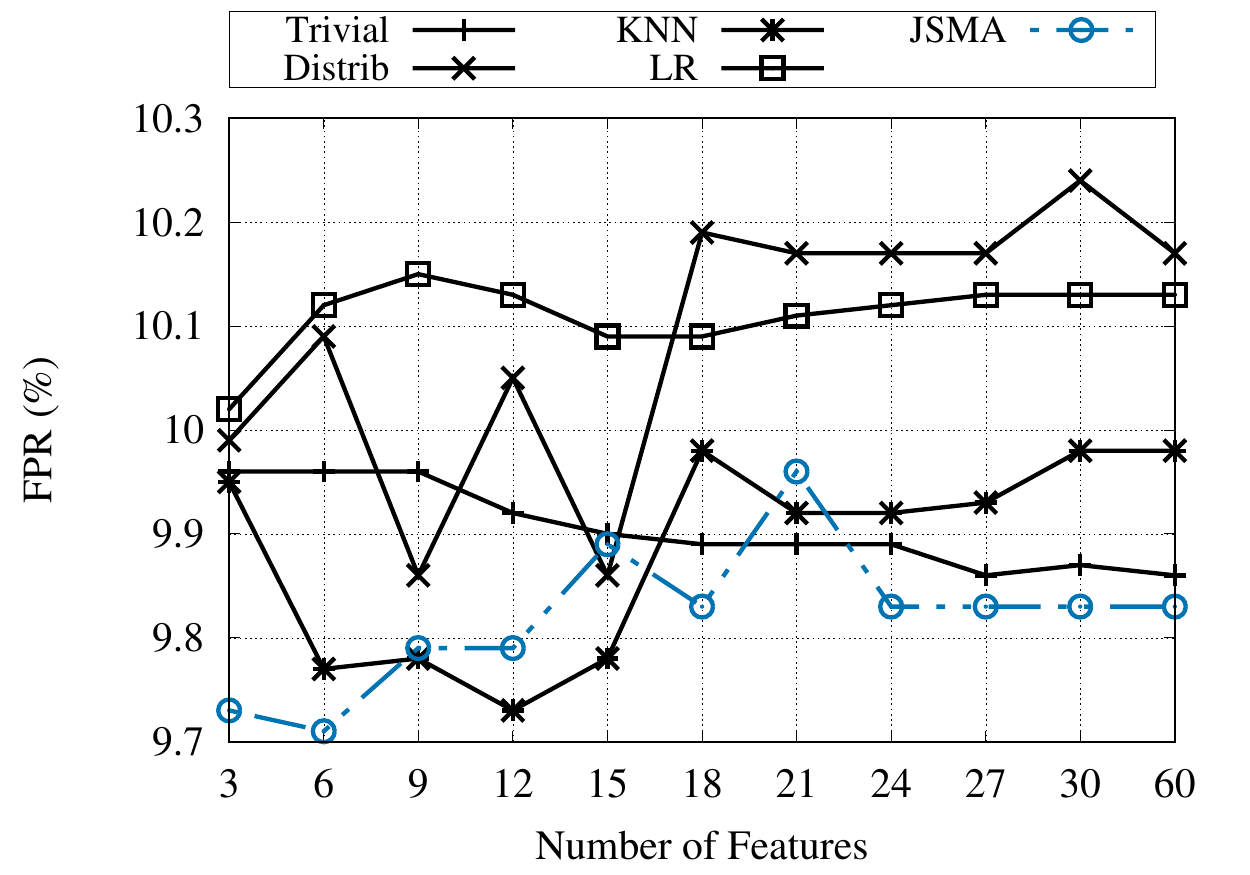}
			\caption{\small Contagio-permission}
			\label{fig:fig4e}
 	\end{subfigure}
   \begin{subfigure}{0.32\textwidth}
 		\centering
 		\includegraphics[width=\linewidth]{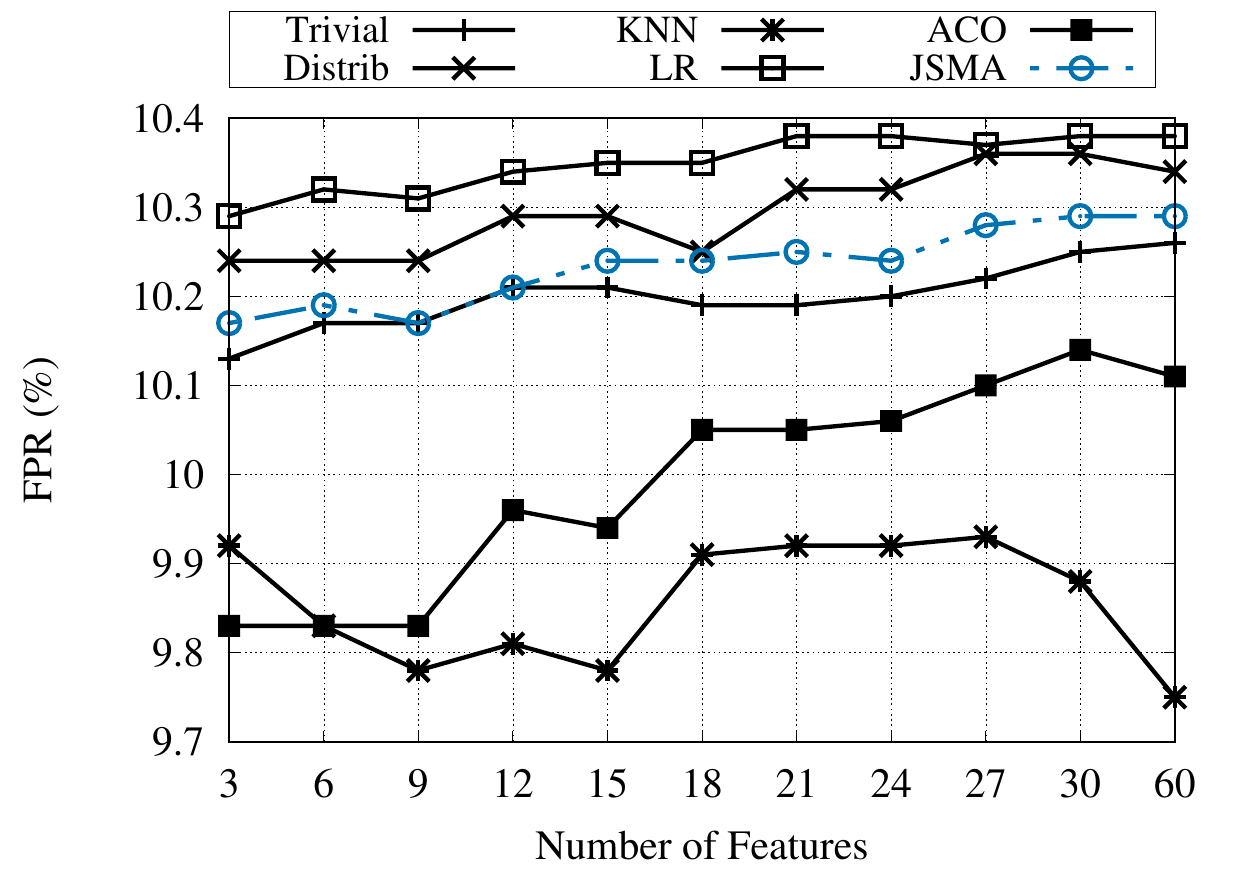}
		\caption{\small Contagio-intent}
			\label{fig:fig4f}
 	\end{subfigure} 
 	\begin{subfigure}{0.32\textwidth}
 		\centering
 		\includegraphics[width=\linewidth]{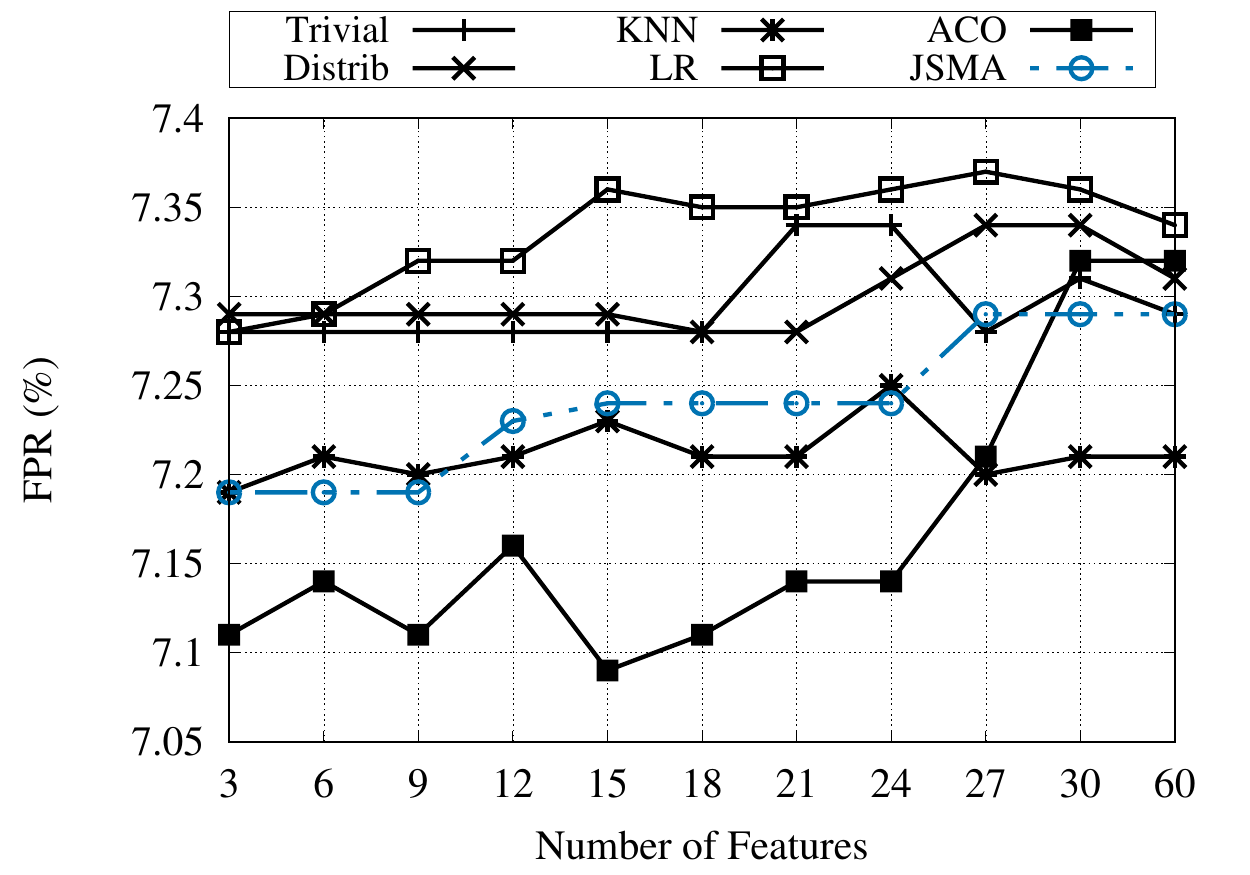}
			\caption{\small Genome-API}
			\label{fig:fig4g}
 	\end{subfigure}  
    \begin{subfigure}{0.32\textwidth}
 		\centering 
 		\includegraphics[width=\linewidth]{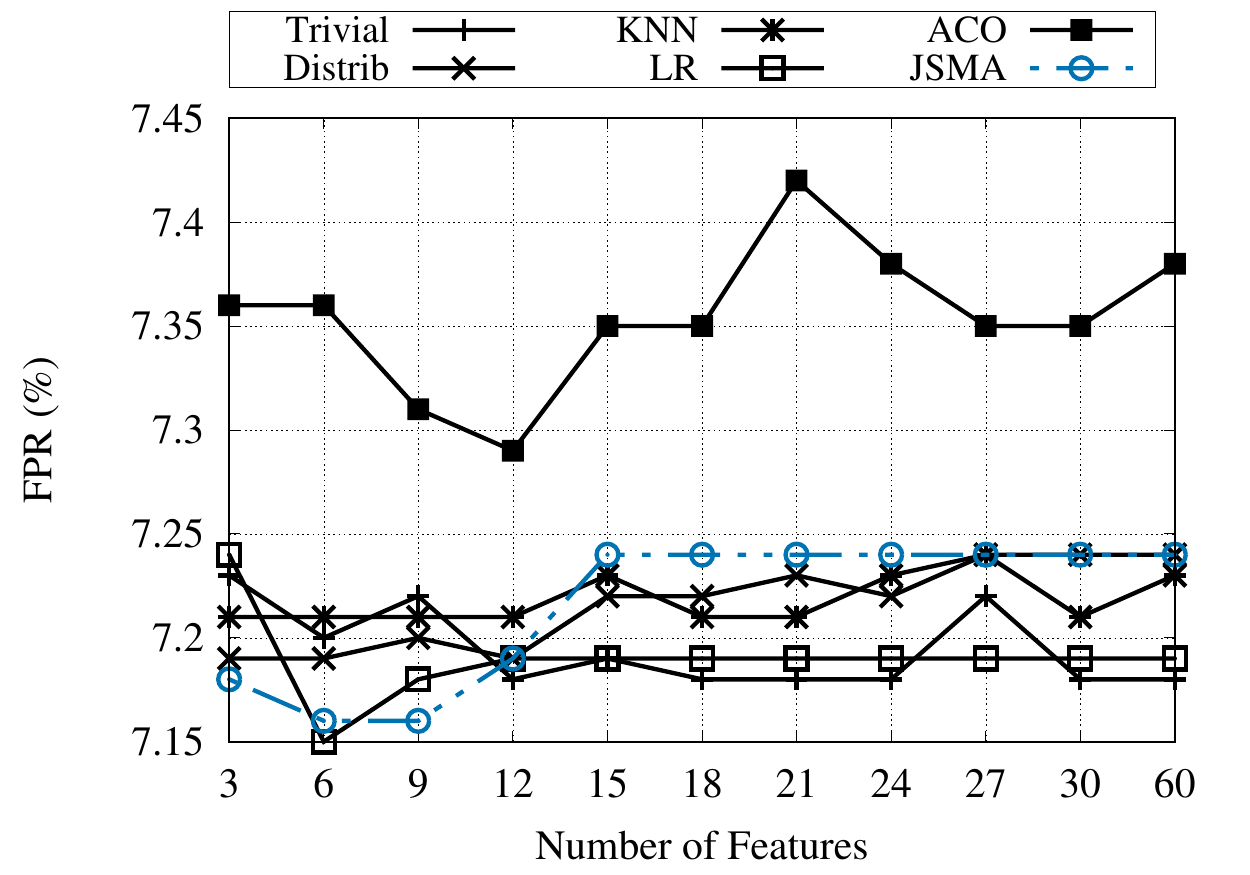}
			\caption{\small Genome-permission}
			\label{fig:fig4h}
 	\end{subfigure}
   \begin{subfigure}{0.32\textwidth}
 		\centering
 		\includegraphics[width=\linewidth]{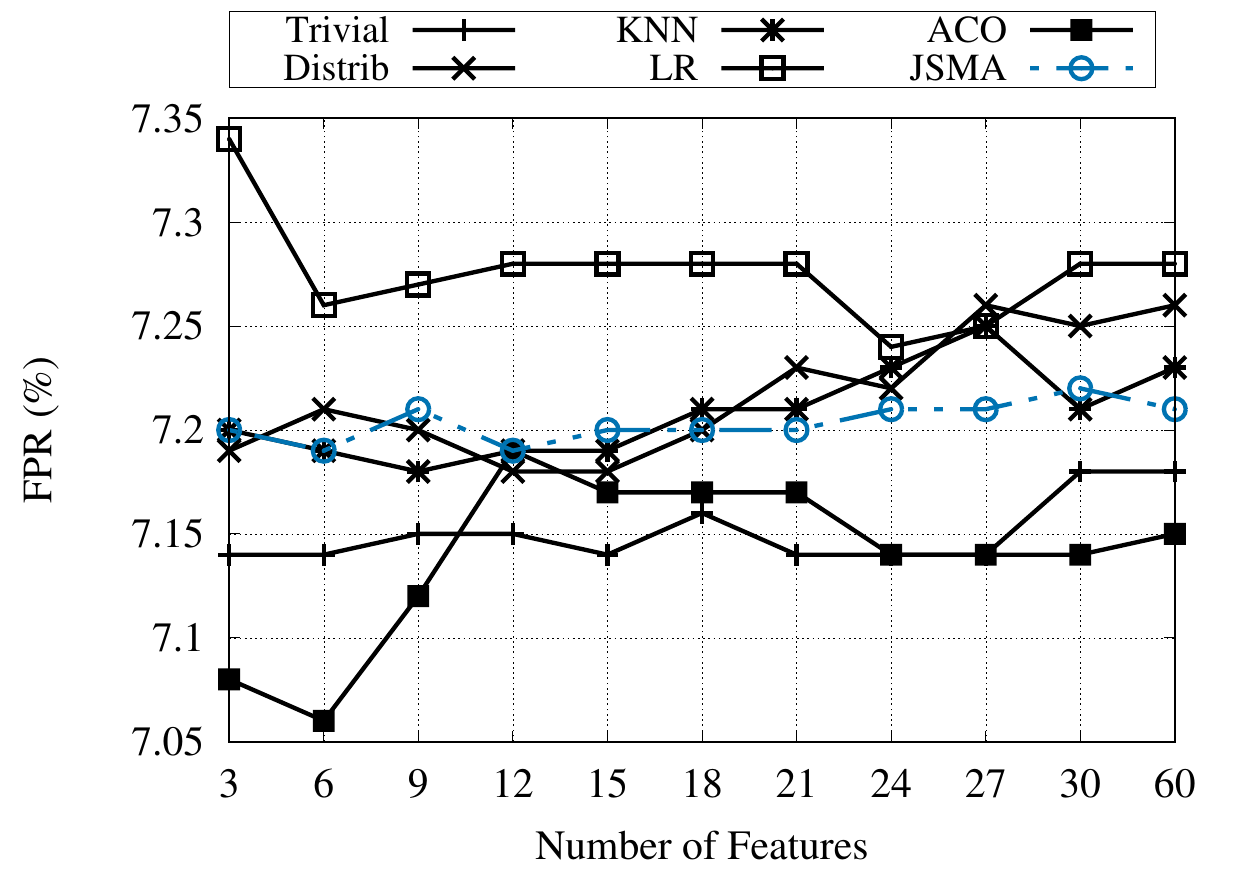}
		\caption{\small Genome-intent}
			\label{fig:fig4i}
 	\end{subfigure} 
 	\hfill
        \caption{\small Comparison between attack algorithms in terms of FPR for various numbers of selected features using an RF classifier for API, intent and permission data.}
			\label{fig:fig4}
		\end{figure*}
\subsubsection{Evaluation of different features}\label{sec:5.b.1}
 In the first results, we present the machine learning metrics for our attack scenarios and defense algorithms versus the JSMA attack method~\cite{grosse2017adversarial} for different numbers of selected features. We present three sets of plots. In the first group, we compare the FPR metric described in Section~\ref{SimulationSetup} for various attack algorithms; in the second group, we compare the aforementioned FPR ratio for the defense algorithms; and in the last group, we validate the AUC metric for the defense and attack algorithms to indicate the success of the proposed attacks and defenses.
 
\paragraph{Comparison of attack algorithms: }\label{sec:5.b.1.1}
 In the following figures, we compare the FPR values for our attack algorithms and the JSMA attack algorithm, using the API and permission data from the Drebin, Contagio and Genome datasets. As we can see, the three sets of comparison plots (Figs.~\ref{fig:fig4a}-\ref{fig:fig4c}, Figs.~\ref{fig:fig4d}-\ref{fig:fig4f}, and Figs.~\ref{fig:fig4g}-\ref{fig:fig4i}) show that for the API and intent type files, the LR attack performs much better than the other algorithms (i.e., for each of the four selected samples, the LR attack can modify the features of one of them in such a way to fool the classifier). This effect can be explained by considering that performing predictive analysis on prominent features avoids overfitting in the poison learning model, which leads to a greater FPR. For permission type apps, the FPR value fluctuates and depends on the features selected. Hence, the FPR values for all attacks do not change smoothly. This value is always higher for the LR, Distribution and KNN attack algorithms than for the JSMA method~\cite{grosse2017adversarial}.
 
 	\begin{figure*}
    \begin{subfigure}{0.32\textwidth}
 		\centering 		\includegraphics[width=1\linewidth]{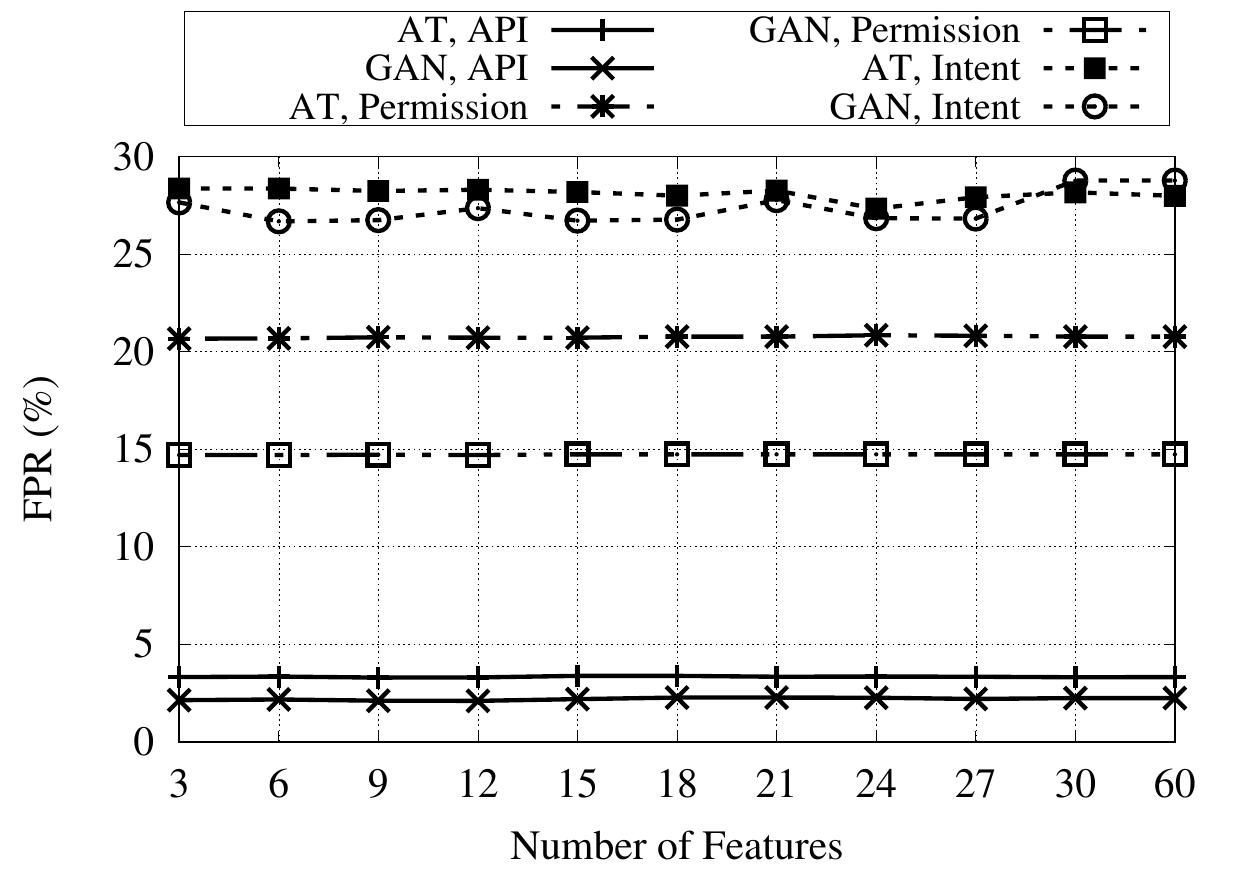}
 		\caption{\small   Drebin}
			\label{fig:fig5a}
 	\end{subfigure} 
   \begin{subfigure}{0.32\textwidth}
 		\centering 		\includegraphics[width=\linewidth]{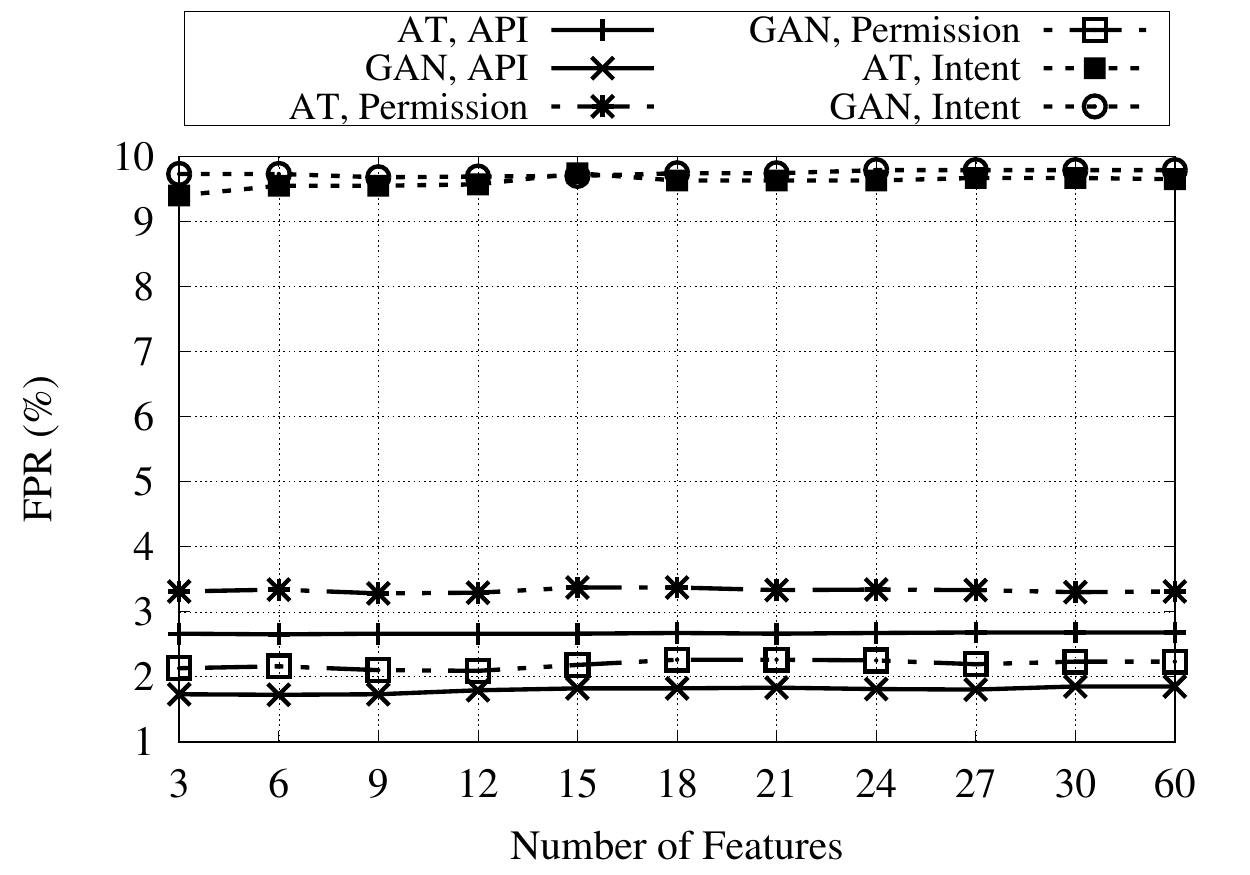}
			\caption{\small   Contagio}
			\label{fig:fig5b}
 	\end{subfigure} 
    \begin{subfigure}{0.32\textwidth}
 		\centering 		\includegraphics[width=\linewidth]{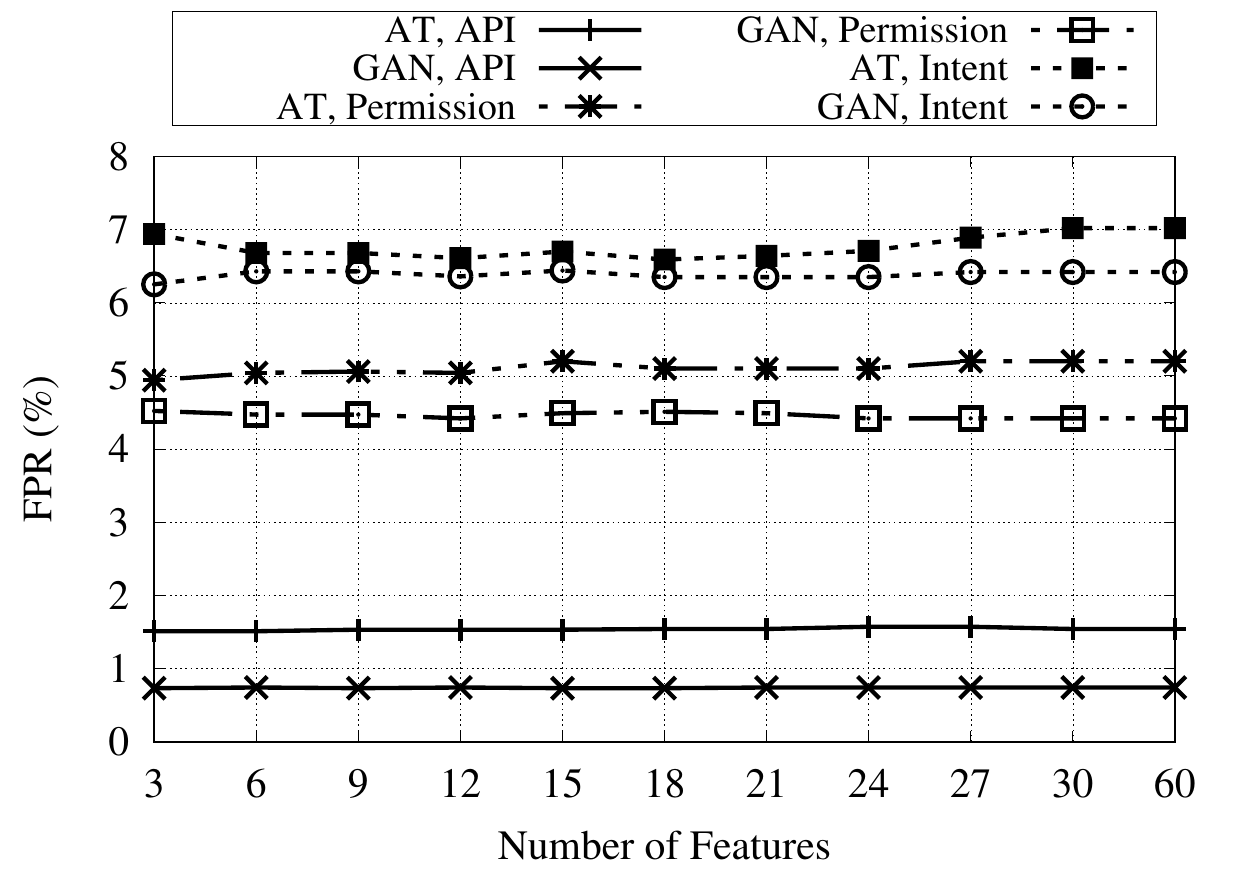}
		\caption{\small   Genome}
			\label{fig:fig5c}
 	\end{subfigure}
        \caption{\small Comparison of defense algorithms in terms of FPR over various numbers of selected features using an RF classifier for API, intent and permission data (AT= adversarial training).}
			\label{fig:fig5}
		\end{figure*}
  \paragraph{Comparison of AUC for different datasets}
	\begin{figure*}[!htb]
    \begin{subfigure}{0.33\textwidth}
 		\centering 		\includegraphics[width=\linewidth]{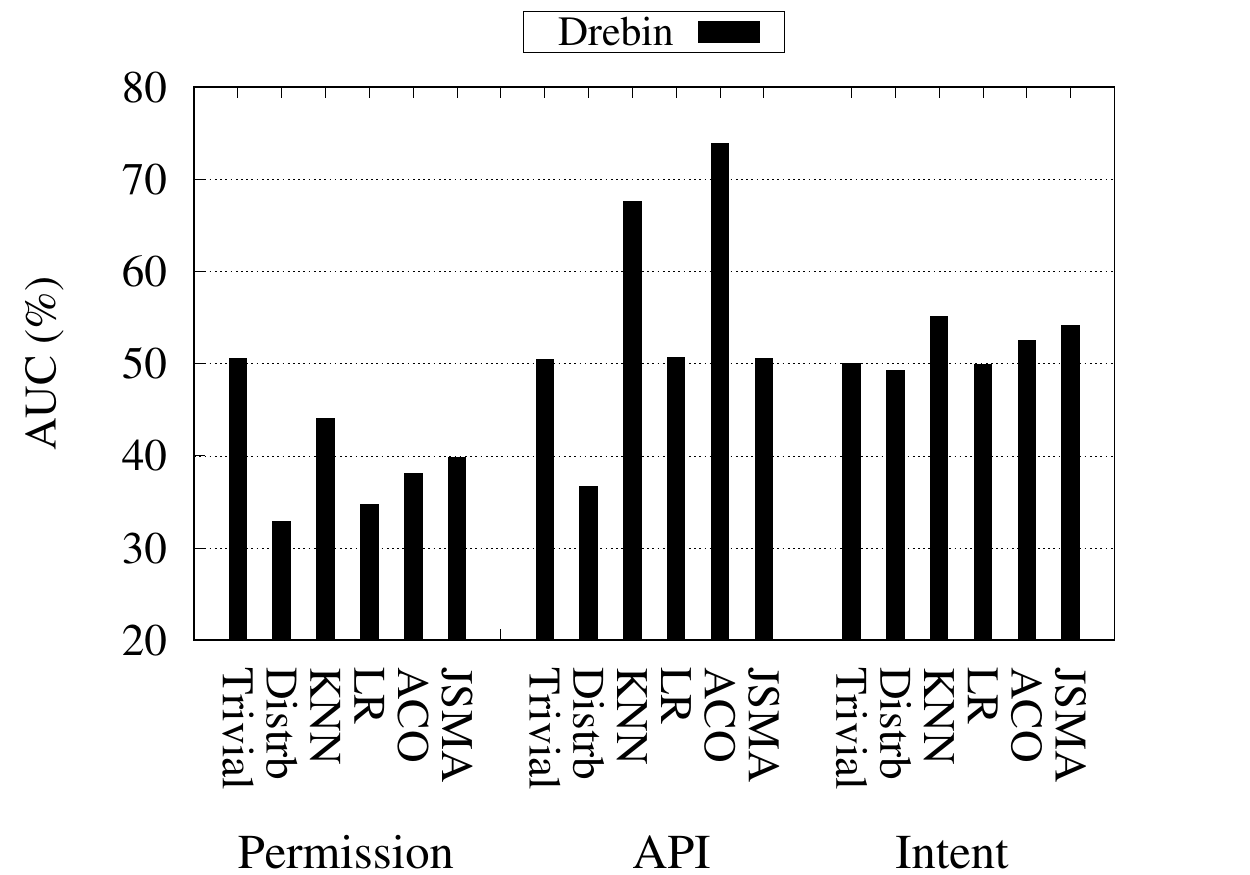}
 		\caption{\small Attack-Drebin}
			\label{fig:fig6a}
 	\end{subfigure} 
   \begin{subfigure}{0.33\textwidth}
 		\centering 		\includegraphics[width=\linewidth]{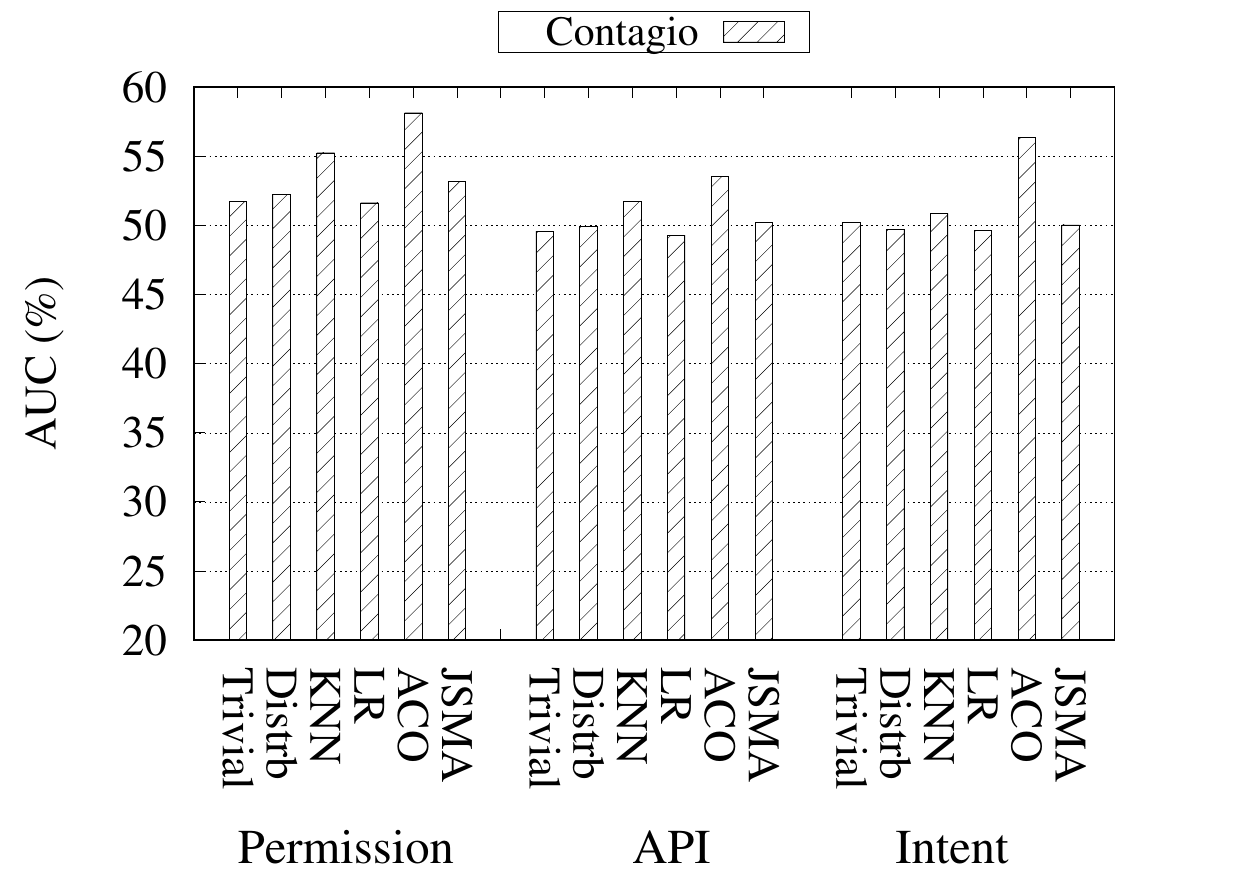}
			\caption{\small Attack-Contagio}
			\label{fig:fig6b}
 	\end{subfigure} 
    \begin{subfigure}{0.33\textwidth}
 		\centering 		\includegraphics[width=\linewidth]{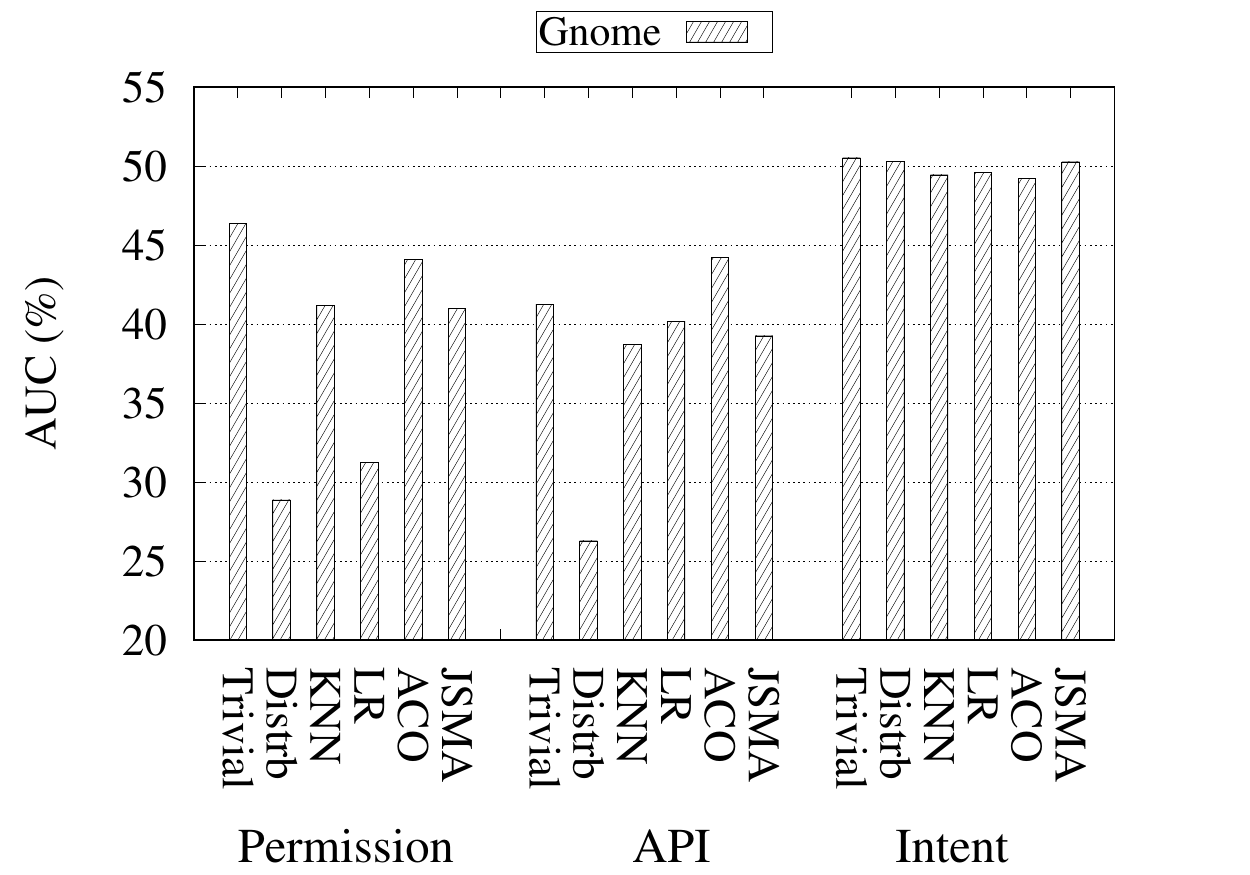}
		\caption{\small Attack-Genome}
			\label{fig:fig6c}
 	\end{subfigure}
 	 \begin{subfigure}{0.33\textwidth}
 		\centering 		\includegraphics[width=\linewidth]{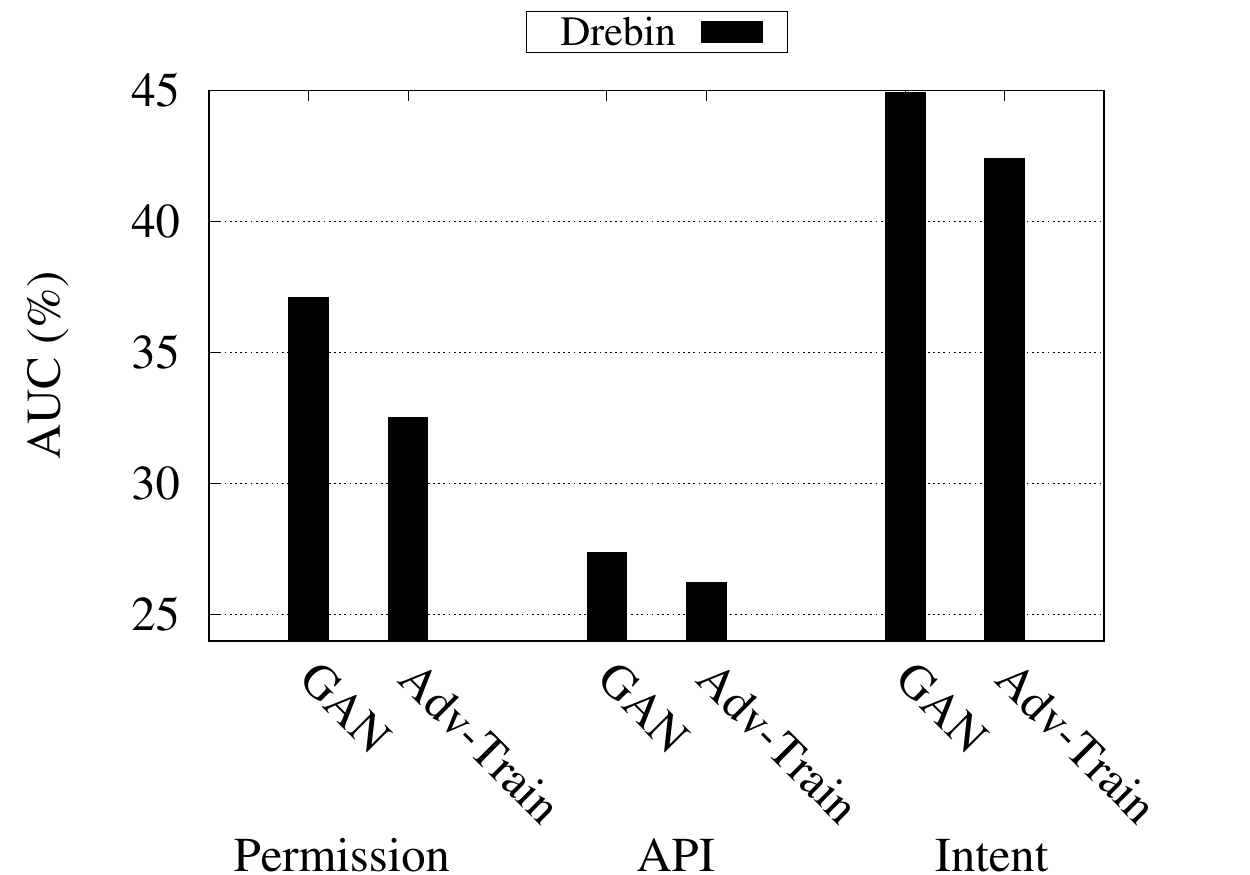}
 		\caption{\small Defense-Drebin}
			\label{fig:fig6d}
 	\end{subfigure} 
   \begin{subfigure}{0.33\textwidth}
 		\centering 		\includegraphics[width=\linewidth]{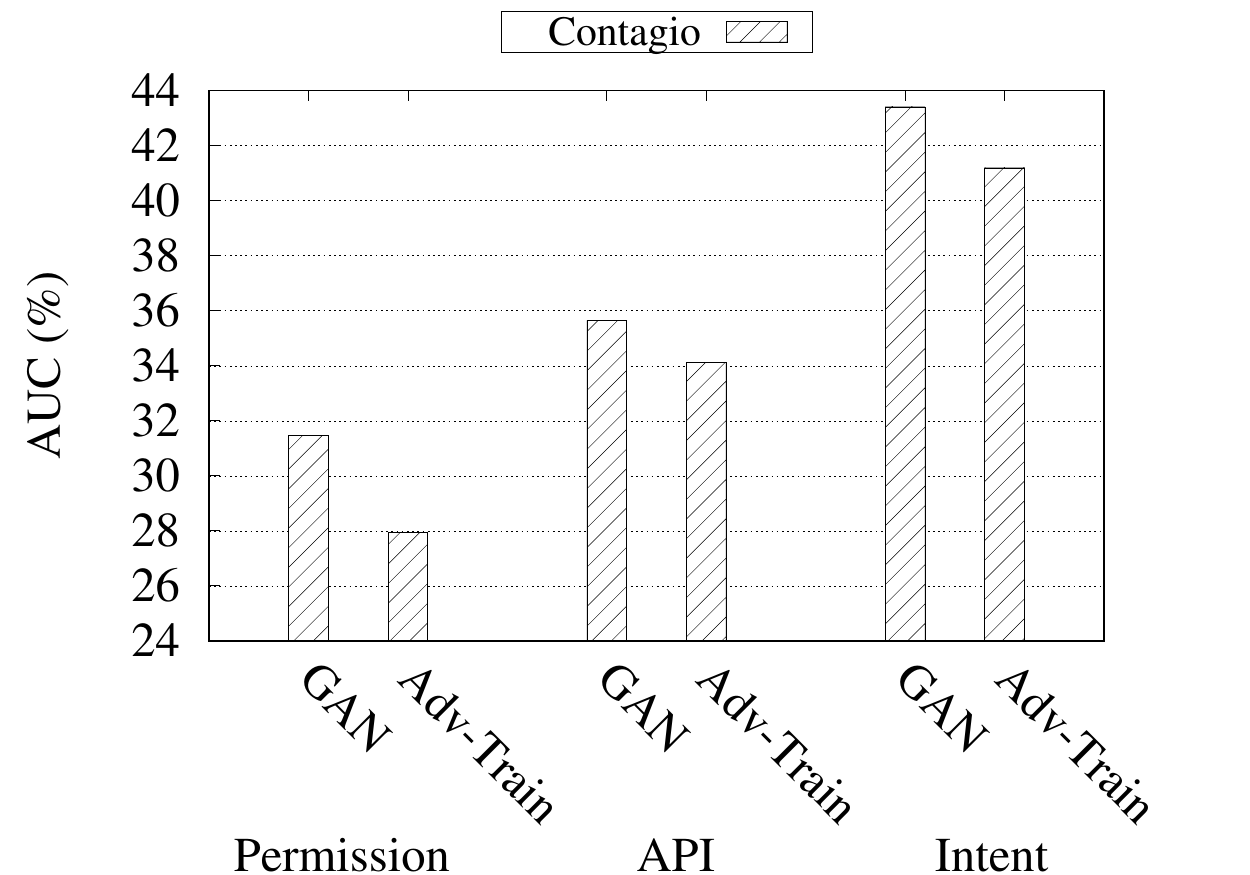}
			\caption{\small Defense-Contagio}
			\label{fig:fig6e}
 	\end{subfigure} 
    \begin{subfigure}{0.33\textwidth}
 		\centering 		\includegraphics[width=\linewidth]{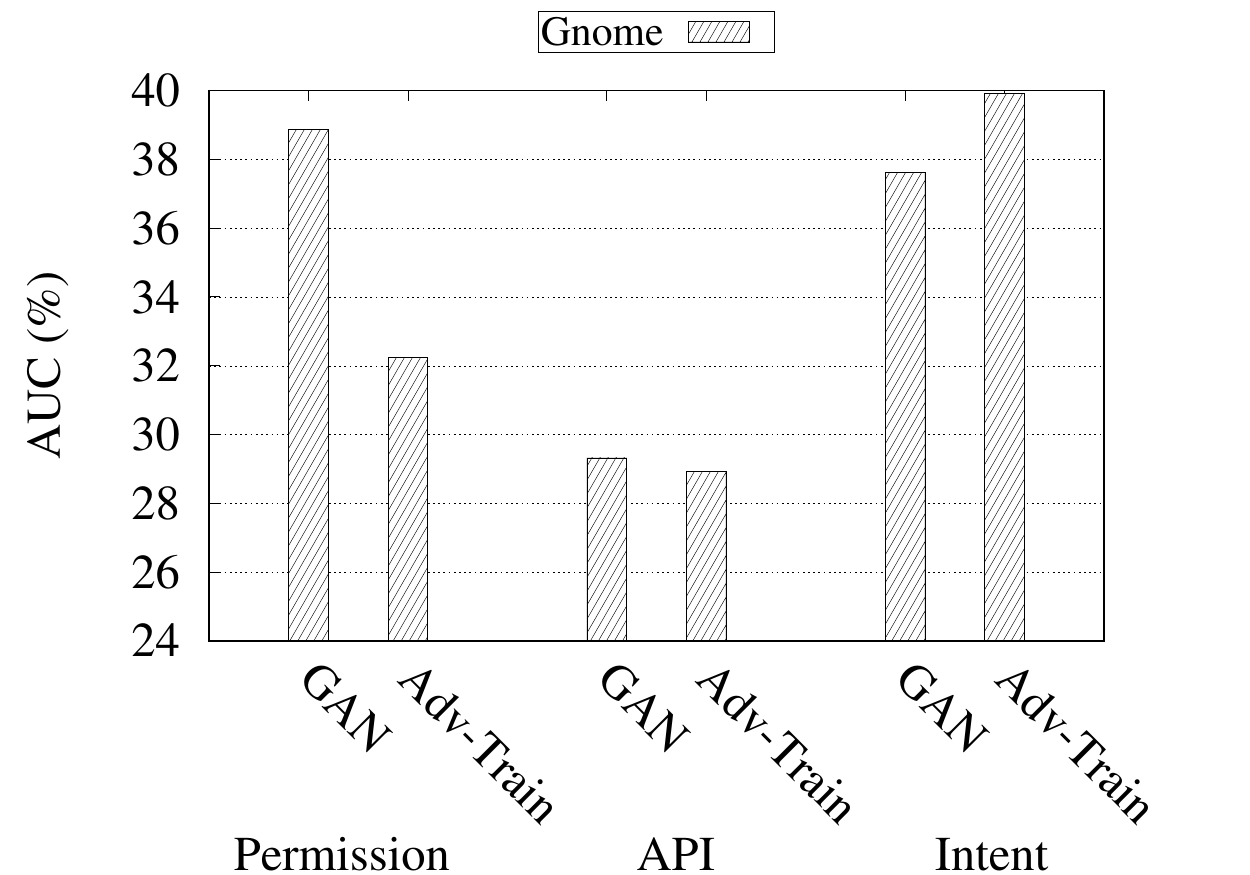}
		\caption{\small Defense-Genome}
			\label{fig:fig6f}
 	\end{subfigure}
        \caption{\small Comparisons of the average AUC value for attack/defense algorithms on different datasets using RF classifier for API, permission, and intent data.}
			\label{fig:fig6}
		\end{figure*}
		
 \paragraph{Comparison of the defense algorithms: }\label{sec:5.b.1.2}
 To fully evaluate our defense algorithms, we test the FPR rate for the various number of feature and file types using various datasets, as shown in Fig.~\ref{fig:fig5}. The following figures show the adversarial training and GAN defense algorithms. From Fig.~\ref{fig:fig5}, we can elicit two main conclusions: firstly, the performance of the GAN defense method (the latter defense method) is better than the former defense algorithm, and the differences between the FPR rates for GAN and the adversarial training algorithm for API files are about 1.1\%, 1\%, and 0.75\% for the Drebin, Contagio and Genome datasets, respectively; and secondly, the differences in the FPR rates for the GAN defense method compared with the LR algorithm for API files (i.e. the highest FPR rate attack algorithm for API files) for the Drebin, Contagio and Genome datasets are about 25\%, 9\%, and 6\%. This indicates that our GAN defense algorithm is an efficient method for reducing the misclassification rate of authentic programs as malware, and is comparable with other results.

In Fig.~\ref{fig:fig6}, we present the AUC values for the attack and defense algorithms. Fig.~\ref{fig:fig6a} shows the AUC values for the attack algorithms (including JSMA~\cite{grosse2017adversarial}) for the permission, API and intent data for the Drebin, Bagging, and Gnome datasets using RF classifier. For an attacker's point of view, it is vital to obtain smaller AUC values that allow adversarial samples to be generated and the learning model to be easily modified. From this figure, we can draw three conclusions. Firstly, attack algorithms perform better for permission data than for API and intent data. This means that our attack algorithms can manipulate the sample data to produce adversarial samples and fool the classification algorithm more easily. Secondly, the LR and distribution attack algorithms have lower values for AUC for all datasets compared with the JSMA algorithm and with respect to the FPR values presented in Fig.~\ref{fig:fig4}. Hence, these two algorithms can be selected as stronger attack methods. Finally, our algorithms show better AUC results for the Drebin dataset for all three file types. This means that the average rate of AUC for all of our attack algorithms applied to permission data from the Drebin dataset is about 15\%, which is 11\% lower than for the API and intent data. However, the average AUC rate for all three file types is approximately the same for the Contagio dataset.   

{A comparison between the defense algorithms presented in Figs.~\ref{fig:fig6d}-\ref{fig:fig6f} shows that it is essential to obtain a significant value of AUC. Hence, from these figures, we can see that the results for the proposed defense methods applied to the API data of the Contagio dataset perform better than for the Drebin and Genome datasets. In contrast, when we perform our defense algorithms on the permission data, the AUC results for the Contagio dataset have smaller values compared with the Drebin and Genome datasets. When we run the defense algorithms on the intent files, the AUC ratio for the Drebin dataset is higher than permission and API file types for the two other datasets. Moreover, the GAN defense method has higher AUC values than the adversarial training defense algorithm, and the rates are approximately the same for both file types. Hence, GAN is an efficient solution that can be applied to all datasets, and especially the Drebin dataset, as a defense against the adversarial example produced by the attack algorithms. As a conclusion, we understand that Drebin dataset can be an efficient dataset that we can perform attack/defense algorithms on it and evaluate our methods compared with the JSMA method~\cite{grosse2017adversarial}.}  

\subsubsection{Performance of attack/defense algorithms on the Drebin dataset}\label{sec:5.b.2}
  
First, we evaluate the average FPR values for all feature lengths for the various attack algorithms, using three types of files for the Drebin dataset and all classifiers, as described in Section~\ref{sec:5.b.2.0}. We then select the much more easily modifiable features, which can accelerate the attack process as described in Section~\ref{sec:5.b.2.1}. Finally, we evaluate the robustness of the attack and defense strategies against the JSMA method~\cite{grosse2017adversarial} and present the ML metrics for the selected features described in Section~\ref{sec:5.b.2.2}.
  
\begin{figure}[!htb]
 		\centering 		\includegraphics[width=\columnwidth]{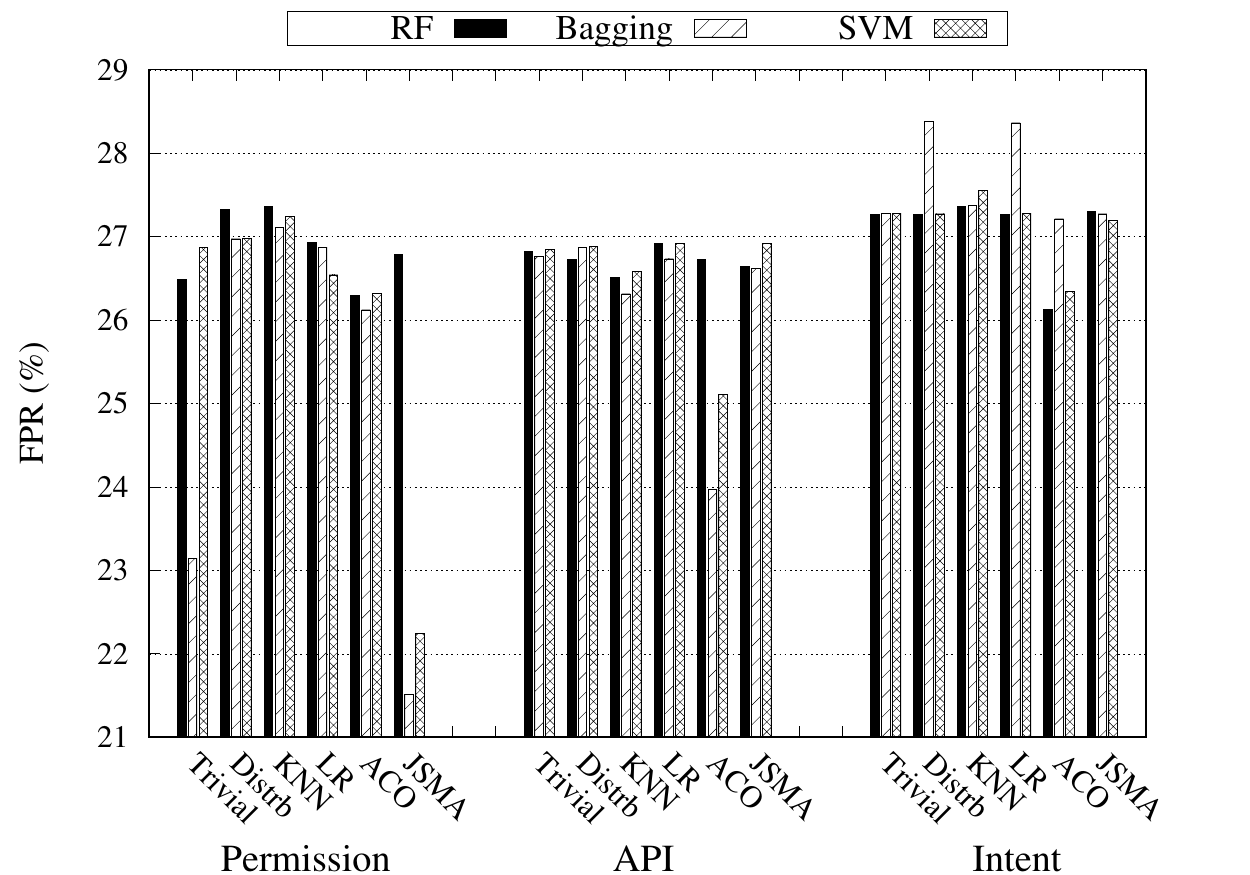}
        \caption{\small   Comparison of average FPR values for attack algorithms for the Drebin dataset using different classifiers and different file types.}
			\label{fig:fig7}
		\end{figure}

  \noindent\paragraph{Performance of FPR vs. file types vs. classifiers}\label{sec:5.b.2.0}
  As an attacker, we need to narrow the attack algorithm targets using the specific classifier and file types. We therefore calculate the FPR value for each file type when manipulated with attack algorithms for various classification methods. Figure~\ref{fig:fig7} demonstrates the results of this evaluation and suggests two findings. Firstly, the average FPR value ratio for the API and permission files is lower than for the intent files. This drives the attack algorithms to concentrate on the last group of files (see the right-hand set of bar charts in Fig.~\ref{fig:fig7}). Hence, the intent file type is much easier to modify using attack algorithms. Secondly, most of our algorithms can produce a larger value of FPR than the JSMA method, for all classification methods. However, the SVM classification algorithm is the weakest classification algorithm, and RF is the strongest. 

  \noindent\paragraph{Performance of reasonable feature selection}\label{sec:5.b.2.1}In this study, an attacker needs to know the minimum number of features of each sample in order to manipulate them and falsify the classifier. We therefore compare and evaluate the FPR rate for different numbers of selected features. Table~\ref{tab2} shows the FPR rate for each attack algorithm for the various numbers of features chosen to run the intent data in the Drebin dataset using the RF classifier. From Table~\ref{tab2}, we can conclude that when increasing the number of features for testing in each scenario, the average FPR remains approximately fixed, while the time of execution clearly increases.

\begin{table*}[!htpb]
\centering
\caption{\small {FPR rate in (\%) for each attack algorithm vs. the number of selected features for permission data in the Drebin dataset using an RF classifier.}}
\label{tab2}
\footnotesize{
\begin{tabular}{|l|p{0.55cm}|p{0.55cm}|p{0.55cm}|p{0.55cm}|p{0.55cm}|p{0.55cm}|p{0.55cm}|p{0.55cm}|p{0.55cm}|p{0.55cm}|p{0.55cm}|}
\hline
\rowcolor{LightCyan} 
&\multicolumn{11}{|c|}{\textbf{Number of selected features}}\\ \cline{3-12}

\rowcolor{LightCyan}
\multirow{-2}{*}{\textbf{Scenario}} &\multicolumn{1}{|c|}{\textbf{3}}&\multicolumn{1}{|c|}{\textbf{6}}&\multicolumn{1}{|c|}{\textbf{9}}&\multicolumn{1}{|c|}{\textbf{12}}&\multicolumn{1}{|c|}{\textbf{15}}&\multicolumn{1}{|c|}{\textbf{18}}&\multicolumn{1}{|c|}{\textbf{21}}&\multicolumn{1}{|c|}{\textbf{24}}&\multicolumn{1}{|c|}{\textbf{27}}&\multicolumn{1}{|c|}{\textbf{30}}&\multicolumn{1}{|c|}{\textbf{60}}\\\hline
\cellcolor[HTML]{E8E8AB}\textbf{Trivial}&\cellcolor{blue!10}27.27&27.39&26.91&27.21&27.22&27.29&27.29&27.29&27.29&27.29&27.29\\\hline
\cellcolor[HTML]{E8E8AB}\textbf{Distrbut.} &\cellcolor{blue!10}27.26&27.27&27.26&27.33&27.36&27.36&27.36&27.40&27.40&27.40&27.40\\\hline
\cellcolor[HTML]{E8E8AB}\textbf{KNN} &\cellcolor{blue!10}27.35&27.35&27.23&27.29&27.38&26.99&27.25&27.32&26.91&26.98&26.98\\\hline
\cellcolor[HTML]{E8E8AB}\textbf{LR} &\cellcolor{blue!10}27.36&27.36&27.31&27.34&27.41&27.43&27.45&27.43&27.43&27.43&27.43\\\hline
\cellcolor[HTML]{E8E8AB}\textbf{ACO} &\cellcolor{blue!10}26.86&26.86&26.89&26.86&26.86&27.04&27.04&27.04&27.08&27.08&27.08\\\hline
\cellcolor[HTML]{E8E8AB}\textbf{JSMA~\cite{grosse2017adversarial}} &\cellcolor{blue!10}27.23&27.24&27.23&27.19&27.26&27.28&27.28&27.30&27.30&27.29&27.30\\\hline
\end{tabular}}
\end{table*}

  \noindent\paragraph{Performance of ML metrics for three selected features}\label{sec:5.b.2.2} 
  In the following, we evaluate the robustness of our classifiers when encountering five different attacks and two defense scenarios for three selected features. Table~\ref{tab3} shows the results of the proposed algorithms for the intent data from the Drebin dataset. As shown in this table, we first apply classification algorithms without an attack strategy. It is noticeable that the accuracy of the classifiers without an attack is above 84\%. The maximum FPR value with no attack is 13.23\% for the RF classifier. Furthermore, the FPR values for the two other classifiers are even less than 13\%.

\begin{table*}
\centering
\caption{\small {Accuracy, FPR, precision and recall values for intent data of the Drebin dataset (Acc= accuracy; Prs= precision; Rec= recall; Distrbut= distribution attack; AT= adversarial training).}}
\label{tab3}
\footnotesize{
\setlength\tabcolsep{1.5pt} 
{\begin{tabular}{|>{\centering\arraybackslash}m{0.9cm}|p{1.75cm}|p{0.65cm}|p{0.65cm}|p{0.65cm}|p{0.5cm}||p{0.75cm}|p{0.65cm}|p{0.65cm}|p{0.65cm}||p{0.75cm}|p{0.65cm}|p{0.65cm}|p{0.65cm}|}
\hline
\rowcolor{LightCyan} 
&&\multicolumn{12}{|c|}{\textbf{Classifiers}}\\ \cline{3-14}

\rowcolor{LightCyan} 
&&\multicolumn{4}{|c|}{\textbf{Random Forest (RF)}}&\multicolumn{4}{|c|}{\textbf{Bagging}}&\multicolumn{4}{|c|}{\textbf{SVM}}\\ \cline{3-13}

\rowcolor{LightCyan}
\multirow{-3}{*}{\begin{sideways}{\textbf{Type}}\end{sideways}}&\multirow{-3}{*}{\textbf{Scenarios.}} &\multicolumn{1}{|c|}{\textbf{Acc.}}&\multicolumn{1}{|c|}{\textbf{FPR}}&\multicolumn{1}{|c|}{\textbf{Prs}}&\multicolumn{1}{|c|}{\textbf{Rec}}&\multicolumn{1}{|c|}{\textbf{Acc}}&\multicolumn{1}{|c|}{\textbf{FPR}}&\multicolumn{1}{|c|}{\textbf{Prs}}&\multicolumn{1}{|c|}{\textbf{Rec}}&\multicolumn{1}{|c|}{\textbf{Acc}}&\multicolumn{1}{|c|}{\textbf{FPR}}&\multicolumn{1}{|c|}{\textbf{Prs}}&\multicolumn{1}{|c|}{\textbf{Rec}}\\\hline\hline%

\cellcolor[HTML]{FAEC0D}\textbf{Norm}&\cellcolor[HTML]{E8E8AB}\textbf{Without}&\cellcolor{blue!10}84.16&\cellcolor{blue!10}13.23&\cellcolor{blue!10}0.91&\cellcolor{blue!10}0.63&\cellcolor{blue!10}80.09&\cellcolor{blue!10}12.99&\cellcolor{blue!10}0.86&\cellcolor{blue!10}0.71&\cellcolor{blue!10}79.76&\cellcolor{blue!10}13.05&\cellcolor{blue!10}0.87&\cellcolor{blue!10}0.67\\\hline\hline

\cellcolor[HTML]{FAEC0D}&\cellcolor[HTML]{E8E8AB}\textbf{Trivial}&72.60&27.27&0.31&0.17&72.05&27.12&0.27&0.16&72.70&27.49&0.31&0.16\\\cline{2-14}

\cellcolor[HTML]{FAEC0D}&\cellcolor[HTML]{E8E8AB}\textbf{Distrbut.}&72.50&27.26&0.26&0.12&\cellcolor[HTML]{FAF0E6}71.02&27.42&0.29&0.14&72.59&27.80&0.28&0.14\\\cline{2-14}

\cellcolor[HTML]{FAEC0D}&\cellcolor[HTML]{E8E8AB}\textbf{KNN}&\cellcolor[HTML]{FAF0E6}71.50&27.35&0.29&0.01&72.4&27.39&0.30&0.07&71.80&27.77&0.29&0.06\\\cline{2-14}

\cellcolor[HTML]{FAEC0D}&\cellcolor[HTML]{E8E8AB}\textbf{LR}&72.50&\cellcolor[HTML]{FAF0E6}27.36&0.34&0.19&71.40&\cellcolor[HTML]{FAF0E6}27.56&0.29&0.11&\cellcolor[HTML]{FAF0E6}71.56&\cellcolor[HTML]{FAF0E6}27.92&0.31&0.16\\\cline{2-14}

\cellcolor[HTML]{FAEC0D}\multirow{-3}{*}{\begin{sideways}{\textbf{Attack}}\end{sideways}}&\cellcolor[HTML]{E8E8AB}\textbf{ACO}&78.86&26.86&0.31&0.21&77.29&26.31&0.29&0.13&72.60&26.14&0.30&0.17\\\cline{2-14}

\cellcolor[HTML]{FAEC0D}&\cellcolor[HTML]{E8E8AB}\textbf{JSMA~\cite{grosse2017adversarial}}&\cellcolor{blue!10}73.21&\cellcolor{blue!10}27.23&\cellcolor{blue!10}0.26&\cellcolor{blue!10}0.16&\cellcolor{blue!10}72.21&\cellcolor{blue!10}27.50&\cellcolor{blue!10}0.25&\cellcolor{blue!10}0.14&\cellcolor{blue!10}72.36&\cellcolor{blue!10}27.21&\cellcolor{blue!10}0.27&\cellcolor{blue!10}0.18\\\hline\hline

\cellcolor[HTML]{FAEC0D}&\cellcolor[HTML]{E8E8AB}\textbf{AT}&79.10&17.53&0.68&0.35&80.11&17.45&0.71&0.39&79.07&17.16&0.69&0.43\\\cline{2-14}

\cellcolor[HTML]{FAEC0D}\multirow{-2}{*}{\begin{sideways}{\textbf{Dfns}}\end{sideways}}&\cellcolor[HTML]{E8E8AB}\textbf{GAN}&\cellcolor[HTML]{FAF0E6}80.93&\cellcolor[HTML]{FAF0E6}14.36&0.82&0.46&\cellcolor[HTML]{FAF0E6}82.54&\cellcolor[HTML]{FAF0E6}15.95&0.79&0.41&\cellcolor[HTML]{FAF0E6}80.54&\cellcolor[HTML]{FAF0E6}15.14&0.85&0.51\\\hline
\end{tabular}}

}
\end{table*}

  The trivial algorithm inserts random noise into samples. Hence, the accuracy value for the trivial algorithm is approximately the same for all classification algorithms and is about 72\%. Conversely, the FPR value increases, but due to the lack of targeted changes in this method, the FPR is lower than for the other proposed attacks. The distributed attack only manipulates the features of the malware samples that are within the distribution of benign samples in the training set. As can be seen from Table~\ref{tab3}, this algorithm is more successful in reducing the accuracy than the other methods for the Bagging classifier and reduces it by about 9\% compared to the rate before the attack. The distribution algorithm also has the lowest precision compared to the other proposed attack scenarios and has a value similar to the JSMA method of about 26\%. Hence, this attack offers a reasonable level of accuracy and can be a suitable option for the Drebin dataset using the Bagging classifier. The KNN attack scenario is an aggressive attack since it reduces the accuracy to the lowest value for the RF classifier compared to the other attack algorithms. This attack selects $k$ benign samples near to each malware sample and changes their features. Hence, a KNN attack can obtain the highest recall values. When using the LR attack scenario as a discriminator, we tried to change the benign samples near to this discriminator by adding features from malware samples to fool the classifier. From the attacker’s point of view, this algorithm is the dominant attack, since it has the lowest value for FPR. The FPR values obtained by applying the LR attack on the RF, Bagging and SVM classifiers are 27.36\%, 27.56\%, and 27.92\%, respectively. Finally, the ACO attack scenario produces adversarial samples using the ant colony optimization algorithm and adds them to the dataset. The results of the experiments show that this method is similar to other methods in terms of the FPR value, but the accuracy value does not change significantly. This means that the level of accuracy is not reduced and the value of the FPR is no higher than for the other attack algorithms.
  
  For the defense algorithms, we expect the accuracy of the classification to increase and the FPR value to decrease. From Table~\ref{tab3}, we observe that the GAN-based method always has higher accuracy than the adversarial training method. The FPR values for the adversarial training method with the RF, Bagging, and SVM classifiers are lower than for the GAN defense method. In Table~\ref{tab3}, we highlight the highest FPR and lowest accuracy values among the attack algorithms and the lowest FPR and highest accuracy among the defense algorithms; these correspond to the best attack and best defense algorithms.


  \subsubsection{Evaluation of detection time}\label{sec:5.b.3}
    In this section, the execution times for the attack algorithms and the defense solutions are compared. Table~\ref{tab4} compares the time required for training, testing, poisoning and defense for the different proposed algorithms. As can be seen from the proposed algorithms, the trivial attack is the fastest attack method, since it randomly selects and modifies the features. The ACO attack algorithm is the slowest attack algorithm to generate adversarial samples. The KNN attack algorithm requires the calculation of the distance between the adjacent samples, and takes much more time than the other attack methods. In terms of the classification algorithms, the RF algorithm time consumption for the training and testing phases is lower than for the Bagging and SVM algorithms. By comparing the proposed algorithms with the JSMA method, we find that the execution time for the proposed methods is better in most cases than for the JSMA algorithm. In terms of the time taken by the attack algorithm on the permission, API and intent data types, it is easy to see that the time required for API data files is larger than for the two other files types, and this rate is lower for the RF classification algorithm. 

    Both of the proposed defense algorithms need a certain amount of time to apply the defense mechanism in the poisoned dataset, which we call the defense period, and this is presented in the same cell as the poisoning time for each classification algorithm. According to the last six rows of Table~\ref{tab4}, the defense time for the adversarial training algorithm is about half of the time taken by the GAN defense learning model against adversarial example injections. However, the duration of the training and testing phases for both defense algorithms are comparable. It is also clear that API apps require more time than other file types for both defense algorithms. Overall, the SVM classifier requires more time than RF for all three phases (i.e., training, testing, and defense). Table~\ref{tab4} confirms that GAN is the best defense algorithm; among the classification algorithms, RF is the most efficient classifier and requires less time than the other classification algorithms.

\begin{table*}[!htpb]
\centering
\caption{\small Execution time in seconds ($s$) for training, testing and application/refining of the poisoning phases on attack/defense algorithms for all file types, datasets and classifications (TRN=training phase; TES=test phase; POS=poisoning phase; AT= adversarial training; DFT= defense time).}
\label{tab4}
\footnotesize{
\begin{tabular}{|>{\centering\arraybackslash}m{0.1cm}|m{1.5cm}|p{1.6cm}|p{0.55cm}|p{0.55cm}|p{0.65cm}||p{0.55cm}|p{0.55cm}|p{0.65cm}||p{0.55cm}|p{0.55cm}|p{0.65cm}|}
\hline
\rowcolor{LightCyan} 
\multicolumn{3}{|c|}{\textbf{Drebin Dataset}}&\multicolumn{9}{|c|}{\textbf{Classifiers}}\\ \hline

\rowcolor{LightCyan} 
\multicolumn{3}{|c|}{\textbf{Time (s)}}&\multicolumn{3}{|c|}{\textbf{RF}}&\multicolumn{3}{|c|}{\textbf{Bagging}}&\multicolumn{3}{|c|}{\textbf{SVM}}\\ \hline

\rowcolor{LightCyan}
\multicolumn{2}{|c|}{\textbf{Scenario}}&\textbf{File Type}&\multicolumn{1}{|c|}{\textbf{TRN}}&\multicolumn{1}{|c|}{\textbf{TES}}&\multicolumn{1}{|c|}{\textbf{POS}}&\multicolumn{1}{|c|}{\textbf{TRN}}&\multicolumn{1}{|c|}{\textbf{TES}}&\multicolumn{1}{|c|}{\textbf{POS}}&\multicolumn{1}{|c|}{\textbf{TRN}}&\multicolumn{1}{|c|}{\textbf{TES}}&\multicolumn{1}{|c|}{\textbf{POS}}\\\hline\hline

\cellcolor[HTML]{c6e2ff}&\cellcolor[HTML]{FAEC0D}&\cellcolor[HTML]{E8E8AB}\textbf{Permission}&3.32&1.48&0.03&2.85&3.50&0.02&3.50&3.56&0.03\\\cline{3-12}
\cellcolor[HTML]{c6e2ff}&\cellcolor[HTML]{FAEC0D}&\cellcolor[HTML]{E8E8AB}\textbf{API}&3.95&2.76&0.22&2.43&3.62&0.28&3.73&4.34&0.218\\\cline{3-12}
\cellcolor[HTML]{c6e2ff}&\cellcolor[HTML]{FAEC0D}\multirow{-3}{*}{\textbf{Trivial}}&\cellcolor[HTML]{E8E8AB}\textbf{Intents}&3.09&1.97&0.02&0.59&4.09&0.15&2.5&3.85&0.96\\\cline{2-12}\cline{2-12}

\cellcolor[HTML]{c6e2ff}&\cellcolor[HTML]{FAEC0D}&\cellcolor[HTML]{E8E8AB}\textbf{Permission}&3.56&1.72&0.10&4.48& 4.78&0.18&3.40&3.86&0.109\\\cline{3-12}
\cellcolor[HTML]{c6e2ff}&\cellcolor[HTML]{FAEC0D}&\cellcolor[HTML]{E8E8AB}\textbf{API}&3.38&3.21&0.36&2.09&3.43&0.37&3.84&3.05&0.296\\\cline{3-12}
\cellcolor[HTML]{c6e2ff}&\cellcolor[HTML]{FAEC0D}\multirow{-3}{*}{\textbf{Distrbut.}}&\cellcolor[HTML]{E8E8AB}\textbf{Intents}&2.82&1.91&0.06&0.57&3.84&0.062&2.42&2.9&0.093\\\cline{2-12}\cline{2-12}

\cellcolor[HTML]{c6e2ff}&\cellcolor[HTML]{FAEC0D}&\cellcolor[HTML]{E8E8AB}\textbf{Permission}&3.51&1.23&1.92&4.20&5.51&2.89&3.82&3.2&1.59\\\cline{3-12}
\cellcolor[HTML]{c6e2ff}&\cellcolor[HTML]{FAEC0D}&\cellcolor[HTML]{E8E8AB}\textbf{API}&3.91&3.22&2.75&2.93&4.79&2.00&3.48&3.84&1.89\\\cline{3-12}
\cellcolor[HTML]{c6e2ff}&\cellcolor[HTML]{FAEC0D}\multirow{-3}{*}{\textbf{KNN}}&\cellcolor[HTML]{E8E8AB}\textbf{Intents}&2.98&1.52&1.86&0.53&4.21&1.50&2.46&3.81&1.48\\\cline{2-12}\cline{2-12}

\cellcolor[HTML]{c6e2ff}&\cellcolor[HTML]{FAEC0D}&\cellcolor[HTML]{E8E8AB}\textbf{Permission}&3.44&2.01&0.13&4.54&5.09&0.26&3.70&3.73&0.01\\\cline{3-12}
\cellcolor[HTML]{c6e2ff}\multirow{-5}{*}{\begin{sideways}{\textbf{Attacks}}\end{sideways}}&\cellcolor[HTML]{FAEC0D}&\cellcolor[HTML]{E8E8AB}\textbf{API}&3.65&2.73&0.90&3.67&4.55&0.93&3.86&3.43&0.1\\\cline{3-12}
\cellcolor[HTML]{c6e2ff}&\cellcolor[HTML]{FAEC0D}\multirow{-3}{*}{\textbf{LR}}&\cellcolor[HTML]{E8E8AB}\textbf{Intents}&2.70&1.27&0.11&0.68&4.52&0.06&3.69&3.82&0.113\\\cline{2-12}\cline{2-12}

\cellcolor[HTML]{c6e2ff}&\cellcolor[HTML]{FAEC0D}&\cellcolor[HTML]{E8E8AB}\textbf{Permission}&4.00&1.96&146.13&4.21&3.04&125.6&3.56&3.53&161.24\\\cline{3-12}
\cellcolor[HTML]{c6e2ff}&\cellcolor[HTML]{FAEC0D}&\cellcolor[HTML]{E8E8AB}\textbf{API}&2.98&3.97&151.76&3.21&4.89&139.13&3.72&3.92&109.6\\\cline{3-12}
\cellcolor[HTML]{c6e2ff}&\cellcolor[HTML]{FAEC0D}\multirow{-3}{*}{\textbf{ACO}}&\cellcolor[HTML]{E8E8AB}\textbf{Intents}&2.79&1.70&143.62&0.62&4.62&99.94&3.61&3.26&176.1\\\cline{2-12}\cline{2-12}

\cellcolor[HTML]{c6e2ff}&\cellcolor[HTML]{FAEC0D}&\cellcolor[HTML]{E8E8AB}\textbf{Permission}&3.70&1.84&0.145&4.34&5.34&0.21&3.98&3.29&0.23\\\cline{3-12}
\cellcolor[HTML]{c6e2ff}&\cellcolor[HTML]{FAEC0D}&\cellcolor[HTML]{E8E8AB}\textbf{API}&3.44&2.81&0.84&3.79&3.74&0.97&3.76&4.16&1.06\\\cline{3-12}
\cellcolor[HTML]{c6e2ff}&\cellcolor[HTML]{FAEC0D}\multirow{-3}{*}{\textbf{JSMA~\cite{grosse2017adversarial}}}&\cellcolor[HTML]{E8E8AB}\textbf{Intents}&2.65&2.34&0.07&0.57&3.89&0.62&2.52&3.29&0.01\\\hline\hline

\rowcolor{LightCyan}
\multicolumn{2}{|c|}{\textbf{Scenario}}&\textbf{File Type}&\multicolumn{1}{|c|}{\textbf{TRN}}&\multicolumn{1}{|c|}{\textbf{TES}}&\multicolumn{1}{|c|}{\textbf{DFT}}&\multicolumn{1}{|c|}{\textbf{TRN}}&\multicolumn{1}{|c|}{\textbf{TES}}&\multicolumn{1}{|c|}{\textbf{DFT}}&\multicolumn{1}{|c|}{\textbf{TRN}}&\multicolumn{1}{|c|}{\textbf{TES}}&\multicolumn{1}{|c|}{\textbf{DFT}}\\\hline\hline 

\cellcolor[HTML]{c6e2ff}&\cellcolor[HTML]{f9f3bd}&\cellcolor[HTML]{E8E8AB}\textbf{Permission}&3.61&1.83&6.99&3.29&1.71&5.51&3.43&3.72&8.20\\\cline{3-12}

\cellcolor[HTML]{c6e2ff}&\cellcolor[HTML]{f9f3bd}&\cellcolor[HTML]{E8E8AB}\textbf{API}&3.18&3.09&	7.75&3.67&3.29&7.64&3.81&3.35&7.44\\\cline{3-12}
\cellcolor[HTML]{c6e2ff}&\cellcolor[HTML]{f9f3bd}\multirow{-3}{*}{\textbf{AT}}&\cellcolor[HTML]{E8E8AB}\textbf{Intents}&2.32&1.89&3.43&2.98&1.94&3.75&2.94&3.03&3.95\\\cline{2-12}\cline{2-12}

\cellcolor[HTML]{c6e2ff}&\cellcolor[HTML]{f9f3bd}&\cellcolor[HTML]{E8E8AB}\textbf{Permission}&3.39&1.61&10.48&3.24&1.83&9.78&3.73&3.27&9.86\\\cline{3-12}
\cellcolor[HTML]{c6e2ff}&\cellcolor[HTML]{f9f3bd}&\cellcolor[HTML]{E8E8AB}\textbf{API}&3.97&3.27&11.24&3.12&2.91&12.43&3.68&3.66&11.89\\\cline{3-12}
\cellcolor[HTML]{c6e2ff}\multirow{-5}{*}{\begin{sideways}{\textbf{Defenses}}\end{sideways}}&\cellcolor[HTML]{f9f3bd}\multirow{-3}{*}{\textbf{GAN}}&\cellcolor[HTML]{E8E8AB}\textbf{Intents}&3.05&1.43&7.78&3.24&1.28&11.4&3.03&3.86&7.90\\\hline
\end{tabular}}
\end{table*}




\section{Discussion and Limitations}\label{discussion}


{The primary goal of crafting adversarial samples is to evaluate the robustness of Machine Learning based malware classifier for identifying malware camouflaged as legitimate samples. In order to ascertain the aforesaid conjecture, we developed malicious samples statistically identical to benign applications by implementing different poisoning methods. In each attack scenario, the functionality of malicious samples was preserved. However structural changes at the Android component level was carried out. To be precise, we considered two restrictions which assure app functionality. Firstly, we modify the manifest features that are related to the AndroidManifest.xml file in any Android app. Secondly, we change the features that are added to the real application. These features are written using a single line of code. Comprehensive experiments carried in this paper clearly depict that malware samples can be shifted to a benign class by altering permission and intents. In the majority of cases, we observed that there was an increase in FPR between 26.5\%-28.5\%. Further, we also conclude that the transformation of a malicious app's to trusted ones could be undertaken with additional efforts, requiring augmentation of large code blocks as compared to attributes like permission and intents. We observed that retraining with adversarial examples with corrected labels and GAN generated samples the trained model can appropriately identify samples drawn from the unseen distribution. One of the main limitations of the proposed method is the detection of malicious samples launching an attack on execution time. This can be addressed by creating a system call flow graph and focusing on critical path depicting frequent operations. Subsequently, feeding the machine learning classifier with the statistics of a sequence of frequent operations. Alternatively, apps, if analyzed independently, may appear legitimate but when they collude shows malicious behavior. Detection of samples exhibiting such behavior is beyond the scope of this work. However, such apps can be detected by estimating flow by representing the information flow using finite state with the output. This way of representing an application state can exhibit fine-grained information flow from multiple states to the subset of states or specific states of given automata. Additionally, we can infer the current state and next state information on an event to discover the vulnerable source and sink pairs.}

\section{Conclusions and Future Work}\label{conclusion}
In this paper, we propose five different attack algorithms: a trivial algorithm, a benign distribution, KNN, LR, and a bio-inspired method based on the ant colony algorithm. We compare these algorithms with the most recent static approach based on a Jacobian method, called JSMA, in terms of providing adversarial examples based on Android mobile data to fool classification algorithms. We also propose two defense algorithms based on adversarial training and GAN architecture. We validate our attack and defense algorithms using three public datasets, namely the Drebin, Genome, and Contagio datasets, using API, intent and permission file types. We test our models before and after implementing attacks on three classification algorithms: the RF, SVM and Bagging algorithms. It is observed that using 300 ranked syntax features of these Android mobile datasets, the benign distribution and LR attack algorithms could fool the classification algorithms using the Drebin dataset. This was particularly true for RF, with an FPR of more than 27\%, an accuracy of less than 72\% and an AUC of up to 50\%. These achievements are interesting as they are higher than the JSMA approach by about 5\% for AUC, 5\% for FPR, and 5\% for accuracy. The GAN approach can also decrease the FPR value by about 10\%, increase the precision by about 50\%, and increase the accuracy by up to 25\%, and can be used as an efficient solution for such attacks.

For the future development of our work, we may further improve the level of per feature robustness against various adversarial manipulations. To do so, we need to add the knowledge of feature robustness to the learning model to make it difficult for the attacker to identify the feature characteristics which are more complicated to manipulate. 

\section{Acknowledgment}
Mauro Conti and Mohammad Shojafar are supported by Marie Curie Fellowships funded by the European Commission (agreement PCIG11-GA-2012-321980) and (agreement MSCA-IF-GF-2019-839255), respectively.
  \bibliographystyle{spmpsci}
  \bibliography{references}

\begin{wrapfigure}{l}{0\textwidth}
{\includegraphics[width=1.15in,height=1.15in,clip,keepaspectratio]{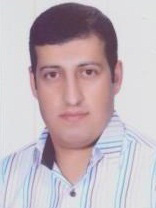}}
\end{wrapfigure}
\noindent\textbf{Rahim~Taheri} received his B.Sc. degree of Computer engineering from Bahonar Technical College of Shiraz and M.Sc. degree of computer networks from Shiraz University of Technology in 2007 and 2015, respectively. Now he is a Ph.D. candidate on Computer Networks in Shiraz University of Technology. In February 2018, he joined to SPRITZ Security \& Privacy Research Group at the University of Padua as a visiting PhD student. His main research interests include machine learning, data mining,  network securities and heuristic algorithms. He currently focused on adversarial machine and deep learning as a new trend in computer security. 

\begin{wrapfigure}[9]{l}{0\textwidth}
{\includegraphics[width=1.15in,height=1.15in,clip,keepaspectratio]{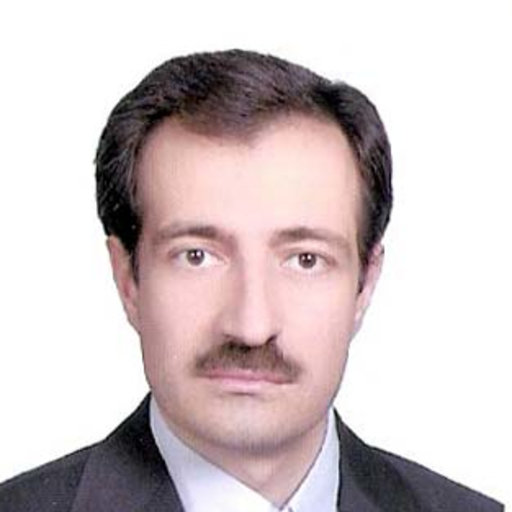}}
\end{wrapfigure}
\noindent\textbf{Reza~Javidan}  received M.Sc. Degree in Computer Engineering (Machine Intelligenceand Robotics) from Shiraz University in 1996. He received Ph.D. degree in Computer Engineering (Artificial Intelligence) from Shiraz University in 2007. Dr. Javidan has many publications in international conferences and journals regarding Image Processing, Underwater Wireless Sensor Networks (UWSNs) and Soft-ware Defined Networks (SDNs). His major fields of interest are Network security, Underwater Wireless Sensor Networks (UWSNs), Software Defined Networks (SDNs), Internet of Things, artificial intelligence, image processing and SONAR systems. Dr. Javidan is an associate professor in Department of Computer Engineering and Information Technology in Shiraz University of Technology.	

\begin{wrapfigure}[9]{l}{0\textwidth}
{\includegraphics[width=1.15in,height=1.15in,clip,keepaspectratio]{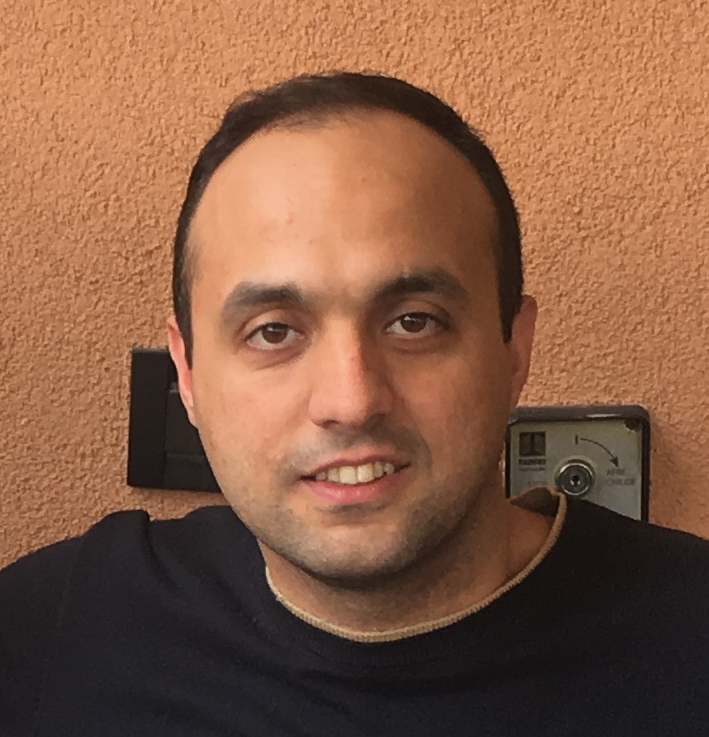}}
\end{wrapfigure}
\noindent\textbf{Mohammad Shojafar} is an Intel Innovator, a Senior IEEE member, a Senior Researcher and a Marie Curie Fellow in the SPRITZ Security and Privacy Research group at the University of Padua, Italy. Also, he was CNIT Senior Researcher at the University of Rome Tor Vergata contributed to 5G PPP European H2020 ``SUPERFLUIDITY'' project and Senior Researcher at the Ryerson University, Toronto, Canada. He is a PI on PRISENODE project, a 275,000 euro Horizon 2020 Marie Curie project in the areas of network security and Fog computing and resource scheduling collaborating between the University of Padua and University of Melbourne. He also was a PI on an Italian SDN security and privacy (60,000 euro) supported by the University of Padua in 2018. He also contributed to some Italian projects in telecommunications like GAUChO— A Green Adaptive Fog Computing and Networking Architecture (400,000 euro), and SAMMClouds- Secure and Adaptive Management of Multi-Clouds (30,000 euro) collaborating among Italian universities. He received a Ph.D. in ICT from Sapienza University of Rome, Italy, in 2016 with an ``Excellent'' degree. His main research interests are in the area of Computer Networks, Network Security, and Privacy. In this area, he published more than 100+ papers in topmost international peer-reviewed journals and conferences, e.g., IEEE TCC, IEEE TNSM, IEEE TGCN, IEEE TSUSC, IEEE Network, IEEE SMC, IEEE PIMRC, and IEEE ICC/GLOBECOM. He served as a PC member of several prestigious conferences, including IEEE INFOCOM Workshops in 2019, IEEE GLOBECOM, IEEE ICC, IEEE UCC, IEEE ScalCom, and IEEE SMC. He was GC in FMEC 2019, INCoS 2019, INCoS 2018, and a Technical Program Chair in IEEE FMEC 2020. He served as an Associate Editor in IEEE Transactions on Consumer Electronics, IET Communication, Springer Cluster Computing, and Ad Hoc \& Sensor Wireless Networks Journals. For additional information: \url{http://mshojafar.com} 

\begin{wrapfigure}[8]{l}{0\textwidth}
{\includegraphics[width=1.15in,height=1.05in,clip,keepaspectratio]{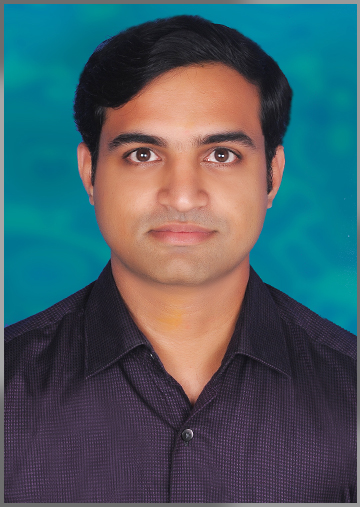}}
\end{wrapfigure}	
\noindent\textbf{Vinod P.} is Post Doc at Department of Mathematics, University of Padua, Italy. He holds his Ph.D in Computer Engineering from Malaviya National Institute of Technology, Jaipur, India. He has more than 70 research articles published in peer reviewed Journals and International Conferences. He is reviewer of number of security journals, and has also served as programme committee member in the International Conferences related to Computer and Information Security. His current research is involved in the development of malware scanner for mobile application using machine learning techniques. Vinod's area of interest is Adversarial Machine Learning, Malware Analysis, Context aware privacy persevering Data Mining, Ethical Hacking and Natural Language Processing.	

\begin{wrapfigure}[9]{l}{0\textwidth}
{\includegraphics[width=1.15in,height=1.15in,clip,keepaspectratio]{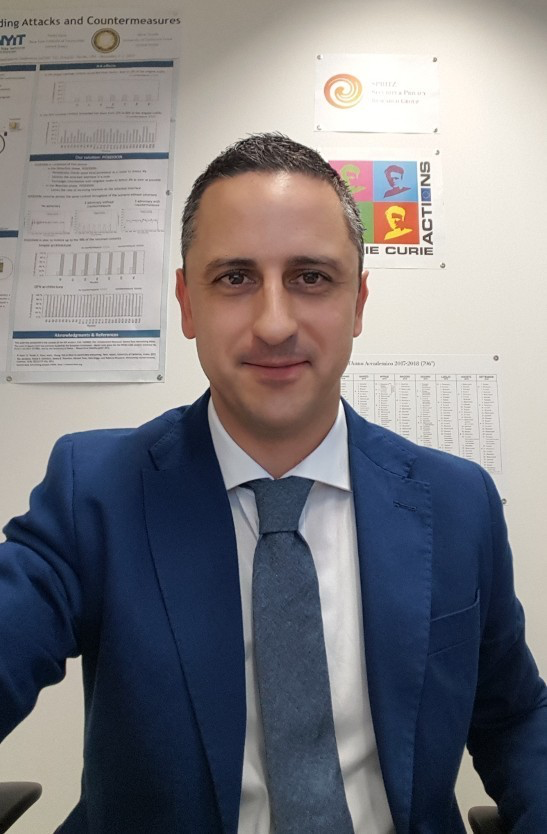}}
\end{wrapfigure}		
\noindent\textbf{Mauro Conti} received his MSc and his PhD in Computer Science from Sapienza University of Rome, Italy in 2005 and 2009. In 2017, he obtained the national habilitation as Full Professor for Computer Science and Computer Engineering. He has been Visiting Researcher at GMU (2008, 2016), UCLA (2010), UCI (2012, 2013, 2014), TU Darmstadt (2013), UF (2015), and FIU (2015, 2016). In 2014, he was elevated to the IEEE Senior Member grade. In 2015 he became Associate Professor, and Full Professor in 2018. He has been awarded with a Marie Curie Fellowship (2012) by the European Commission, and with a Fellowship by the German DAAD (2013). His main research interest is in the area of security and privacy. In this area, he published more than 250 papers in topmost international peer-reviewed journals and conference. He is Associate Editor for several journals, including IEEE Communications Surveys \& Tutorials, IEEE Transactions on Network and Service Management, and IEEE Transactions on Information Forensics and Security. He was Program Chair for TRUST 2015, ICISS 2016, WiSec 2017, and General Chair for SecureComm 2012 and ACM SACMAT 2013. He is Senior Member of the IEEE. For additional information: \url{ http://www.math.unipd.it/~conti/} 

\setcounter{figure}{0} \renewcommand{\thefigure}{A.\arabic{figure}}
\renewcommand*\thetable{A.\arabic{table}}

\section*{Appendix}
{In this section, we use 60 ranked (ascending priorities) dimensional feature vectors out of 300 available features for the study for different datasets. For the sake of simplicity, we present these features for the Drebin dataset (breakdowns are shown in Table~\ref{taba1}). For example, we select and adopt 20 first features of the Drebin dataset for 6\% of selected feature rates.}

\newpage

\begin{sidewaystable}[!htpb]
\centering
\caption{\small {60 ranked features for API, permission and intent in Drebin dataset for training classifiers.}}
\label{taba1}
\resizebox{\textwidth}{!}{
\begin{tabular}{|l|l|l|l|}
\hline

\rowcolor{LightCyan} 
\textbf{Priority}&\textbf{API}&\textbf{Permission}&\textbf{Intent}\\\hline
 1&Landroid/app/WallpaperManager;$->$clear&st.brothas.mtgoxwidget.permission.C2D\_MESSAGE&com.airpush.android.PushServiceStart58925\\\hline
 2&Landroid/content/pm/PackageManager;$->$clearPackagePreferredActivities&st.veezie.full.permission.C2D\_MESSAGE&com.airpush.android.PushServiceStart52518\\\hline
 3&Landroid/content/ContentResolver;$->$getIsSyncable&suedtirolnews.app.permission.C2D\_MESSAGE&com.airpush.android.PushServiceStart12256\\\hline
 4&Landroid/accounts/AccountManager;$->$peekAuthToken&sva.permission.READ_CONTACTS&android.intent.action.ADS\\\hline
 5&Landroid/content/ContentResolver;$->$isSyncPending&sweet.selfie.lite.permission.C2D\_MESSAGE&widget\_pre\_mini\\\hline
 6&Landroid/content/pm/PackageManager;$->$setApplicationEnabledSetting&sworkitapp.sworkit.com.permission.C2D\_MESSAGE&com.airpush.android.PushServiceStart53179\\\hline
 7&Ljava/lang/String;$->$getBytes&tarot.love.money.career.permission.C2D\_MESSAGE&action\_start\_tb\_orderlist\\\hline
 8&Landroid/content/pm/PackageManager;$->$getPackageSizeInfo&taubeta_242.permission.C2D\_MESSAGE&com.amazon.mp3.playbackcomplete\\\hline
 9&Ljava/lang/String;$->$valueOf&taxi.android.client.permission.C2D\_MESSAGE&works.jubilee.timetree.widget.ACTION\_UPDATED\\\hline
 10&Landroid/content/pm/PackageManager;$->$addPreferredActivity&team.rss.liberoquotidiano.permission.C2D\_MESSAGE&com.airasia.mobile.MESSAGE\\\hline
 11&Landroid/widget/QuickContactBadge;$->$assignContactFromEmail&team.vc.gazzettamezzogiorno.permission.C2D\_MESSAGE&ba.app.zbirkaaforizama.ui.RedirectToPlayActivity\\\hline
 12&Landroid/content/pm/PackageManager;$->$getInstalledPackages&th3.toki.MAgames.permission.C2D\_MESSAGE&android.search.action.GLOBAL\_SEARCH\\\hline
 13&Landroid/net/wifi/WifiManager;$->$isWifiEnabled&thecrush.apuld.permission.C2D\_MESSAGE&android.intent.action.payment\\\hline
 14&Ljava/lang/StringBuilder;$->$append&team.vc.liberoedicola.permission.C2D\_MESSAGE&android.appwidget.action.APPWIGET\_ENABLED\\\hline
 15&Landroid/os/Handler;$->$sendMessage&team.vc.sesaab.permission.C2D\_MESSAGE&com.aichess.alarm.message\\\hline
 16&Landroid/provider/Browser;$->$clearSearches&team.vc.umbria.permission.C2D\_MESSAGE&co.kukurin.worldscope.widgets.FavWidgetProvider.NEXT\\\hline
 17&Landroid/app/backup/BackupManager;$->$requestRestore&teatroverdi_11312.permission.C2D\_MESSAGE&widget\_data\_fetching\_completed\\\hline
 18&Landroid/provider/ContactsContract\$Contacts;$->$lookupContact&tech.azhar.livetalk.permission.C2D\_MESSAGE&ReminderReceiver\\\hline
 19&Ljava/lang/StringBuffer;$->$append&tech.nazca.nuvaringreminder.permission.C2D\_MESSAGE&COM\_TAOBAO\_TAE\_SDK\_TRADE\_WEB\_VIEW\_ACTION\\\hline
 20&Landroid/content/ContentResolver;$->$query&telecom.mdesk.permission.WRITE\_SETTINGS&widget.update.time.tick\\\hline
21& Landroid/app/backup/BackupManager;$->$dataChanged&com.peel.lgtv.permission.C2D\_MESSAGE&com.wave.keyboard.prefs\\\hline
22& Landroid/telephony/SmsManager;$->$sendDataMessage&com.pearsoned.aqua.search.SEARCH_PERMISSION&com.webex.meeting.widget.GET\_MEETING\\\hline
23& Landroid/provider/Contacts\$People;$->$addToGroup&com.peaklens.ar.permission.C2D\_MESSAGE&com.soundcloud.android.actions.upload\\\hline
24& Lorg/w3c/dom/Document;$->$getDocumentElement&com.pdfviewer.pdfreader.permission.C2D\_MESSAGE&goAboutActivity\\\hline
25& Landroid/media/MediaPlayer;$->$stop&com.paypal.android.p2pmobile.permission.C2D\_MESSAGE&mmt.intent.action.OFFERS\_DATA\\\hline
26& Landroid/content/ContentResolver;$->$setMasterSyncAutomatically&com.paylink.westernunion.permission.C2D\_MESSAGE&com.tbig.playerprotrial.metachanged\\\hline
27& Landroid/location/LocationManager;$->$setTestProviderEnabled&com.paperlit.android.iltirrenonz.permission.C2D\_MESSAGE&com.tomtop.shop.intent.action.COMMAND\\\hline
28& Landroid/location/LocationManager;$->$setTestProviderLocation&com.paperlit.android.glamouritalia.permission.C2D\_MESSAGE&com.webex.meeting.widget.AUTO\_REFRESH\\\hline
29& Landroid/accounts/AccountManager;$->$setUserData&com.paperlit.android.aditalia.permission.C2D\_MESSAGE&com.wavesecure.show\_upsell\_notification\\\hline
30& Landroid/net/wifi/WifiManager;$->$reconnect&com.panasonic.avc.cng.imageapp.permission.C2D\_MESSAGE&com.voxel.ACTION\_VIEW\_ENTITY\\\hline
31& Ljava/lang/StringBuilder;$->$indexOf&com.parkeon.whoosh.permission.C2D\_MESSAGE&mokeeobm.wktlmyld.mffesfsStart76\\\hline
32& Landroid/telephony/gsm/SmsManager;$->$getDefault&com.parclick.permission.C2D\_MESSAGE&rcs.intent.action.disableRcs\\\hline
33& Ljava/lang/StringBuffer;$->$indexOf&com.parallel.space.lite.helper&com.tsf.shell.themes\\\hline
34& Landroid/content/ContentResolver;$->$addPeriodicSync&com.sec.print.mobileprint.permission.C2D\_MESSAGE&com.sygic.speedcamapp.routechanged\\\hline
35& Landroid/provider/Telephony\$Sms;$->$moveMessageToFolder&com.sec.permission.OTG_CHARGE_BLOCK&com.sygic.aura.ACTION\_DISMISSED\_PROMO\_NOTIFICATION\\\hline
36& Landroid/hardware/Camera;$->$open&com.sec.permission.ACCESSBILITY\_SHARING&com.tipsmusically.music.modules.podcast.player.ACTION\_PLAY\\\hline
37& Landroid/net/wifi/WifiManager;$->$pingSupplicant&com.sec.orca.remoteshare.permission.WRITE\_MEDIA&com.wavesecure.activities.ProgressDialogActivity\\\hline
38& Landroid/provider/ContactsContract\$Contacts;$->$getLookupUri&com.sec.enterprise.mdm.permission.BROWSER_PROXY&com.wikia.discussions.ui.EDIT\_POST\\\hline
39& Landroid/bluetooth/BluetoothAdapter;$->$getName&com.sec.enterprise.knox.permission.KNOX_RESTRICTION&mokeeobm.wktlmyld.mffesfsStart76\\\hline
40& Landroid/net/wifi/WifiManager\$WifiLock;$->$acquire&com.sec.enterprise.knox.permission.KNOX_ATTESTATION&com.vtcreator.android360.notification.PlaceActivity\\\hline
41& Landroid/net/wifi/WifiManager;$->$addNetwork&com.tensportstvlive.org.permission.C2D\_MESSAGE&com.taobao.ugc.mini.service.publish\\\hline
42& Landroid/media/AudioManager;$->$stopBluetoothSco&com.tencent.qqhead.permission.getheadresp&it.libemax.timbrature.DID\_MONITORING\\\hline
43& Landroid/net/ConnectivityManager;$->$requestRouteToHost&com.tencent.photos.permission.DATA&com.soundcloud.android.playback.external.pause\\\hline
44& Landroid/media/AsyncPlayer;$->$stop&com.tencent.news.push.permission.MESSAGE&com.steelstudio.freemusic.musicplayer.update\\\hline
45& Ljava/io/File;$->$delete&com.tencent.news.permission.MIPUSH\_RECEIVE&com.sand.airdroid.action.servers.stop\_all\\\hline
46& Landroid/net/ConnectivityManager;$->$getActiveNetworkInfo&com.tencent.music.data.permission2&com.sand.airdroid.action.app\_manage\_task\_exec\\\hline
47& Landroid/accounts/AccountManager;$->$removeAccount&com.tencent.mm.permission.MM\_MESSAGE&com.esdk.android.intent.THROTTLE\\\hline
48& Landroid/net/ConnectivityManager;$->$setNetworkPreference&com.tencent.mm.location.permission.SEND_VIEW&com.esdk.android.intent.WIC\_POSITION\\\hline
49& Landroid/location/LocationManager;$->$getLastKnownLocation&com.tencent.mm.ext.permission.READ&com.duolingo.action.REMIND\_LATER\\\hline
50& Landroid/app/Instrumentation;$->$sendCharacterSync&com.tencent.mtt.permission.MIPUSH_RECEIVE&com.didi.virtualapk.about\\\hline
51& Landroid/provider/Settings\$Secure;$->$setLocationProviderEnabled&com.tencent.msg.permission.pushnotify&com.antutu.benchmark.marooned.ERROR\\\hline
52& Landroid/bluetooth/BluetoothDevice;$->$createRfcommSocketToServiceRecord&com.tencent.msf.permission.ACCOUNT_NOTICE&com.anndconsulting.mancala.SoundManager\\\hline
53& Ljavax/crypto/Cipher;$->$getInstance&com.tencent.mm.plugin.permission.WRITE&com.apalon.coloring_book.open.ACTION\_DAILY\_PIC\\\hline
54& Landroid/app/Activity;$->$startActivity&com.tencent.mm.plugin.permission.RECEIVE&com.anglelabs.alarmclock.free.act\_timer\_times\_up\\\hline
55& Landroid/os/PowerManager\$WakeLock;$->$release&com.telos.app.im.permission.C2D\_MESSAGE&\\\hline
56& Ljava/lang/Runtime;$->$loadLibrary&com.teletrader.android.permission.C2D\_MESSAGE&com.android.calendar.ACTION\_CELL\_ZERO\_REFRESH\\\hline
57& Landroid/media/MediaRecorder;$->$setAudioSource&com.telekom.tolino.ACCESS_DATA&com.amazon.INSTALL\_REFERRER\\\hline
58& Landroid/app/WallpaperManager;$->$setResource&com.rosegal.permission.C2D\_MESSAGE&com.android.camera.action.CROP\\\hline
59& Landroid/app/Activity;$->$sendOrderedBroadcast&com.roomster.permission.C2D\_MESSAGE&android.nfc.action.TAG\_DISCOVERED\\\hline
60& Landroid/app/Application;$->$sendStickyBroadcast&com.rometic.trulychinese.permission.C2D\_MESSAGE&com.amazon.avod.LOCATION\\\hline
\end{tabular}
}
\end{sidewaystable}
\end{document}